\documentclass[iop]{emulateapj}
\usepackage{graphicx}
\usepackage{apjfonts}
\usepackage{amsmath} 
\usepackage{natbib}
\usepackage{color}
\usepackage[normalem]{ulem}
\newcommand{\prima}{^{\prime}}
\newcommand{\nprima}{n^{\prime}}
\newcommand{\nprimac}{n^{\prime 2}}
\newcommand{\fdeg}{.\!\! ^{\circ}}
\definecolor{al}{rgb}{0.9,0.1,0.0}

\def\be{\begin{equation}}
\def\ee{\end{equation}}
\def\bea{\begin{eqnarray}}
\def\eea{\end{eqnarray}}
\def\Etilde{\tilde{\textbf {E}}^{(\text{t})}}
\def\degree{^{\circ}}
\def\Et{\textbf {E}^{(\text{t})}}
\def\dx{{\rm d}x}

\def\dr{{\rm d}r}
\def\de{{\rm d}\epsilon}
\def\dS{{\rm d}S}

\def\PSF{{\cal S}}
\slugcomment{Submitted to ApJS}
\shorttitle{I. The isotropic case}
\shortauthors{F.J. Bail\'en et al.}

\begin{document}
	
	\title{On Fabry-P\'erot etalon based instruments \\
	I. The isotropic case}
	
	\author{F.J. Bail\'en, D. Orozco Su\'arez, and J.C. del Toro Iniesta}
	\affil{Instituto de Astrof\'isica de Andaluc\'ia (CSIC), Apdo. de Correos 3004, E-18080 Granada, Spain}
	\email{fbailen@iaa.es, orozco@iaa.es, jti@iaa.es}

\begin{abstract}

Here we assess the spectral and imaging properties of Fabry-P\'erot etalons when located in solar magnetographs. We discuss the chosen configuration (collimated or telecentric) for both ideal and real cases. For the real cases, we focus on the implications caused by the polychromatic illumination of the filter, by irregularities presented in the optical thickness of the etalon and by deviations from the ideal illumination in both setups. We first review the general properties of Fabry-P\'erots and then address the different sources of degradation of the spectral transmission profile. We review and extend the general treatment of defects followed by different authors. We discuss the differences between the point-spread-functions of the collimated and telecentric configurations for both monochromatic and (real) quasi-monochromatic illumination of the etalon. The PSF corresponding to collimated mounts show to have a better performance, although varies from point-to-point due to an apodization of the image inherent to this configuration, contrarily to the (perfect) telecentric case, where the PSF remains constant but produce artificial velocities and magnetic field signals because of its strong spectral dependence.  We find that the unavoidable presence of imperfections in the telecentrism produce a decrease of flux of photons and a shift, a broadening and a loss of symmetrization of both the spectral and PSF profiles over the field-of-view, thus compromising their advantages over the collimated configuration. We evaluate these effects for different apertures of the incident beam. 

\end{abstract}


\section{Introduction}\label{sec:general}
Fabry-P\'erot interferometers (filters or etalons) are extensively employed as tunable  monochromators in post-focus astronomical instrumentation. Some examples are the Italian Panoramic Monochromator at THEMIS \citep[and references therein]{themis}, the TESOS spectrometer at the VTT \citep{kentischer}, the Interferometric Bidimensional Spectrometer at the Dunn Solar Telescope of the Sacramento Peak Observatory \citep{cavallini}, the CRisp Imaging SpectroPolarimeter instrument at the Swedish 1-m Solar Telescope \citep{crisp}, the IMaX instrument aboard {\sc Sunrise} \citep{imax},  the GFPI at GREGOR \citep{gregor}, or the PHI instrument on board the Solar Orbiter mission \citep{sophi}. Their main advantage over single-slit based spectrographs is that they allow for fast imaging of the solar scene and for post-facto imaging reconstruction techniques.\footnote{Techniques for imaging reconstruction in spectrographs that employ slits are still at an early stage of development \citep[e.g.,][]{quintero}} They are also preferred against other devices such as Michelson interferometers or Lyot filters in terms of weight and simplicity. When used in combination with a polarimeter, they enable dual-beam polarimetry, which gets rid of the undesired seeing-induced or \emph{jitter}-induced contamination between Stokes parameters. They present, however, both spectroscopic and imaging drawbacks that restrict their performance.

Fabry-P\'erot etalons present a spectral transmission profile characterized by periodic and narrow resonances at certain wavelengths. The position and width of these depend on intrinsic parameters of the etalon, such as its thickness or its refraction index, as well as on the way the filter is illuminated. In particular, the transmission peaks shift towards the blue when the incident angle is different from zero, which implies a variation on the transmission at monochromatic wavelengths. On the other hand, the width of the resonances broaden when imperfections (defects) appear in the etalon, thus degrading the spectral resolution of the filter \citep[e.g.,][]{chabbal,meaburn,hernandez,sloggett}. Departure from collimated illumination (i.e., when the incident beam has a finite aperture) also widens the peaks and shifts them towards shorter wavelengths \citep[e.g.,][]{sloggett,ref:atherton}. Analytical expressions for determining the broadening of the spectral resolution are usually restricted to particular cases, though \citep[e.g., the \emph{limiting} finesse of ][]{chabbal}. In addition, their derivation is sometimes unclear \citep[e.g., the \emph{aperture} finesse of][]{ref:atherton} and the way different defects are added has been subject to debate \citep{sloggett}. We believe that this topic should be revisited in order to clarify the possible discrepancies and to discuss the validity of the expressions given by different authors.

Concerning its imaging properties, Fabry-P\'erots are used in both collimated \citep[e.g.,][]{bendling,imax} and telecentric configurations \citep[e.g.][]{kentischer,sophi}. In the first case, the etalon is located in a pupil plane, so different incidence angles in the etalon are mapped to different pixels of the detector. This means that, in case of a uniform object
field, the image shows different peak intensities
across the detector at monochromatic wavelengths due to the shift induced by the different
incident angles on the etalon over the field of view (FOV). In the (image-space) telecentric configuration, the etalon is located at a focal plane while the exit pupil is located at infinity. In this setup, if perfect, each point of the etalon receives the same cone of rays from the pupil and the passband is kept constant along the FOV. On the other hand, each point of the etalon ``sees'' the pupil as if it was not evenly illuminated. This effect is due to the variation on the incidence angle for rays coming from different parts of the pupil, an effect known as \emph{pupil apodization} that produces variations of the spatial point spread function (PSF) of the system and of the spectral passband across the detector when defects are present in the etalon.

The image degradation introduced by the Fabry-P\'erot in telecentric mode through pupil apodization was evaluated for the first time by \cite{ref:beckers}, who concluded that collimated illumination of the etalon is preferred over the telecentric configuration in diffraction-limited imaging telescopes. The PSF varies from one wavelength to another in the telecentric configuration, which gives raise to artificial line of sight (LOS) velocity signals that may not be corrected for during data pre-processing. Spurious signals on the magnetic field can also appear. The magnitude of these effects will be discussed in a third part of this series of papers. Although his conclusions were valid, \cite{ref:beckers} calculations were not strictly correct as he considered variations in the magnitude of the electromagnetic field but omitted \emph{phase errors}, i.e., fluctuations in the optical phase produced by the multiple reflections of light within the etalon. 
 These fluctuations were incorporated by \cite{vonderluhe}, who concluded that image degradation effects appearing in telecentric configuration are even more pronounced than those predicted by \cite{ref:beckers}. 
According to them, most wavefront degradation comes from pupil apodization instead of from phase fluctuations. \cite{ref:Scharmer} showed that phase variations can be compensated partially by refocusing the instrument as they depend quadratically with the pupil radial coordinate, in the same fashion as a defocus term.

The collimated configuration is not exempt of problems in terms of image degradation either, as substrate surface roughness are amplified due to the high-reflectivity of the etalon surfaces \citep{vonderluhe}. Both amplitude and phase fluctuations in collimated configuration coming from these irregularities were also studied by \cite{ref:Scharmer}, who pointed out that the effects are less strong than predicted by \cite{vonderluhe} but still important, specially for high reflecting etalons. 
Both works suggest, in contrast to \cite{ref:beckers}, that the telecentric configuration is preferred over the collimated one if high image quality is aimed to be achieved. In our opinion, a comparison needs to be revisited. On the one hand, the \cite{vonderluhe} results about the expected wavefront distortion in  a collimated setup look too pessimistic. On the other hand, the arguments by \cite{ref:Scharmer} image degradation in collimated configurations invite to such an in-depth study.

From our point of view several aspects are yet to be studied. First, some of the analytical approximations of the spectral performance of the etalon are not presented within the realm of a consistent theoretical framework and differ from one author to other \citep{sloggett}. Some of them have not been generalized to crystalline etalons (e.g., the \emph{aperture finesse} defined by \citeauthor{ref:atherton} \citeyear{ref:atherton}). Second, the effects of imperfect telecentrism (i.e, of having non-symmetric pupil apodization over the FOV when the exit pupil is not exactly at infinity, such as in real instruments) have not been thoroughly considered yet up to our knowledge. And third, disagreement between authors makes unclear which configuration is to be preferred in terms of both image quality and spectral transmission. In particular, in an imperfect telecentric setup both the PSF and the spectral profile can broaden and become asymmetric over the FOV (see Section \ref{sec:imperfect}). This means, among other things, that the PSF varies from pixel to pixel even if no defects are present in telecentric mode, which can be critical when referring to image quality. Moreover, a spectral shift is also produced over the FOV, so the passband does not remain constant and the advantage of using a telecentric setup is no longer obvious.

On the other hand, etalons are sometimes made up of electro-optical and piezo-electrical crystals for tuning purposes, specially in space applications \citep{imax,sophi}. The tuning is carried out through variations in the refraction index and thickness when applying a voltage. These crystals usually present birefringent properties and, as they are employed in polarimeters, can disturb the polarization properties of the incoming light and corrupt the polarimetric measurement. Anisotropic effects have only been taken into account through numerical experimentation \citep[e.g.,][]{doerr} and will be studied analytically in the second part of this series of papers for both the collimated and telecentric configuration, in terms of spectral and imaging performance.

Here, we first summarize the relevant theory for analyzing the spectroscopic properties of Fabry-P\'erot etalons (Sections \ref{sec:basicparameters} and \ref{sec:transmission}). We then overview the most common optical configurations (Section \ref{sec:configurations}) making emphasis on the possible sources of the spectral profile degradation. We latter analyze the PSF deterioration in both perfect (Section \ref{sec:psf}) and imperfect (Section \ref{sec:imperfect}) telecentric configurations.

\section{Basic parameters and nomenclature}
\label{sec:basicparameters}

A Fabry-P\'erot etalon is nothing but a resonant optical cavity made up of two semi-reflective and semi-transparent surfaces that separate two different optical media of refractive indices $n$ (the external) and $n\prima$ (the internal). Note that single refractive indices implicitly indicate that the media are assumed to be isotropic. Besides, we shall assume that the media are homogeneous.\footnote{An isotropic medium has the same properties and behavior no matter the direction of the light traveling through it because it is characterized by {\em scalar} dielectric permittivity, magnetic permeability, and electrical conductivity. If those physical quantities have no directional variations across the medium, then it is said homogeneous.} These are correct assumptions for, e.g., air-gapped etalons but they are not for crystalline ones. We shall nevertheless keep the assumptions throughout this paper and defer the discussion of anisotropic etalons to the second paper in this series. 

\begin{figure}[b]
	\begin{center}
		\includegraphics[width=0.48\textwidth]{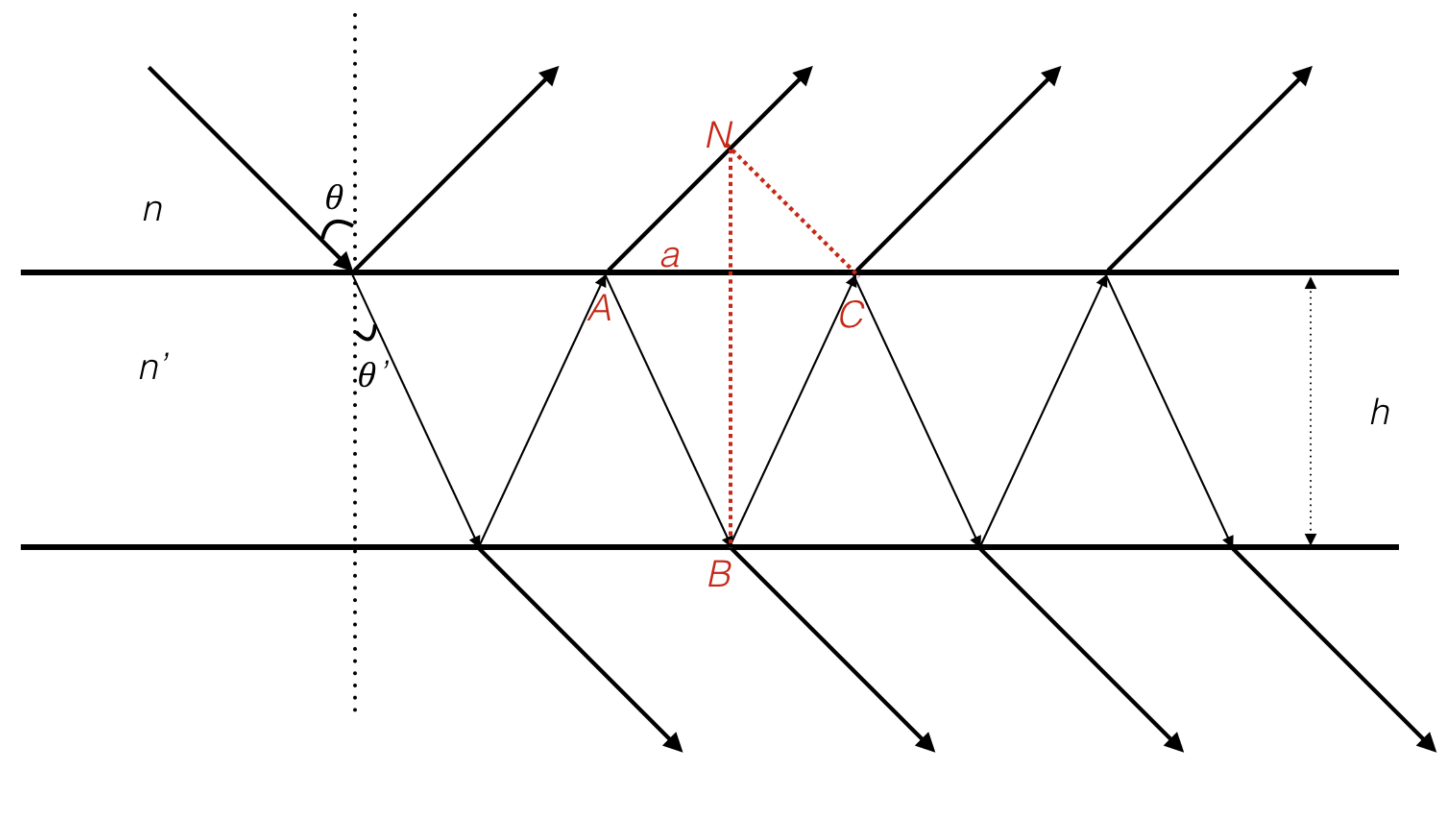}
	\end{center}
	\caption{Transmission and reflection of a plane wave through an isotropic etalon.}
	\label{fig:Etalontransref}
\end{figure}

Such an optical  cavity is also characterized by its geometrical thickness $h$ and by the amplitude reflection and transmission coefficients $r$, $r\prima$, $t$, and $t\prima$ for the external (unprimed) and internal (primed) faces of each surface. As shown in Fig.\ \ref{fig:Etalontransref}, a plane wave impinging the first (top) surface at an angle of incidence $\theta$ partially reflects on and refracts through both surfaces several times. The refraction angle is called $\theta\prima$. 

The fraction of energy reflected from and transmitted through the etalon is given by
\begin{equation}
\label{eq:reflectivity}
R\equiv r^{2} = r^{\prime 2},
\end{equation}
\begin{equation}
\label{eq:transmittivity}
T\equiv tt\prima,
\end{equation}
where we have assumed that $r = -r\prima$. $R$ and $T$ are called the reflectivity and transmittivity of the etalon. If there is no absorption, then 
\begin{equation}
\label{eq:rplust}
R+T=1.
\end{equation}
If, on the contrary, the etalon is coated with a metal that absorbs light with an absorptivity $A$, then
\begin{equation}
\label{eq:rplustplusa}
R+T+A=1.
\end{equation}

\section{The transmission profile for an incident plane wave}
\label{sec:transmission}

Each of the transmitted and reflected rays in Fig.\ \ref{fig:Etalontransref} has a constant phase difference with its predecessor. Let us focus, for instance, in the second and third reflected rays. The optical path difference between them is 
\begin{equation}
\label{eq:path}
\Delta s = n^{\prime} (\overline{AB} + \overline{BC}) - n \, \overline{AN}.
\end{equation}

Since
\begin{equation}
\label{eq:AB}
\overline{AB} = \overline{BC} = \frac{h}{\cos \theta^{\prime}}, 
\end{equation}
\begin{equation}
\label{eq:AC}
\frac{1}{2} \frac{\overline{AC}}{h} = \tan \theta^{\prime},
\end{equation}
and Snell's law,
\begin{equation}
\label{eq:snell}
n \, \sin \theta = n^{\prime} \sin \theta^{\prime},
\end{equation}
one can finally obtain that
\begin{equation}
\label{eq:difpath}
\Delta s = 2 n^{\prime} h \cos \theta^{\prime}.
\end{equation}

The corresponding phase difference between the two rays is 
\begin{equation}
\label{eq:phasedif}
\delta = \frac{4\pi}{\lambda} n^{\prime} h \cos \theta^{\prime} + 2 \phi,
\end{equation}
where $\phi$ is the eventual phase shift introduced by the internal reflections. If the internal surfaces are not coated ---as in crystalline etalons---, then $\phi$ can only be 0 or $\pi$. On the other hand, if the reflecting surfaces are made of metallic films, $\phi$ can take any value in the range $[0,\pi]$ depending on the incident angle. However, if $\theta'$ is close to zero, $\phi$ may be considered to be constant. Furthermore, in general, $h$ is very large compared to $\lambda$. In any case, $\phi$ can be neglected \citep{Hecht}.

According to, e.g., \cite{ref:born}, the ratio between the transmitted, $I^{\rm{(t)}}$, and the incident, $I^{\rm{(i)}}$, intensities can be written as
\begin{equation}
\label{eq:transratio}
g = \frac{I^{\rm{(t)}}}{I^{\rm{(i)}}} = \frac{\tau}{1 + F \sin^{2}(\delta/2)},
\end{equation}
where $\tau$ is the transmission (intensity) factor for normal incidence as given by 
\begin{equation}
\label{eq:tau}
\tau = \left( 1 - \frac{A}{1-R} \right)^{2}
\end{equation}
and parameter $F$ is defined by
\begin{equation}
\label{eq:ef}
F \equiv \frac{4R}{(1-R)^{2}}.
\end{equation}

\begin{figure}
	\begin{center}
		\includegraphics[width=0.48\textwidth]{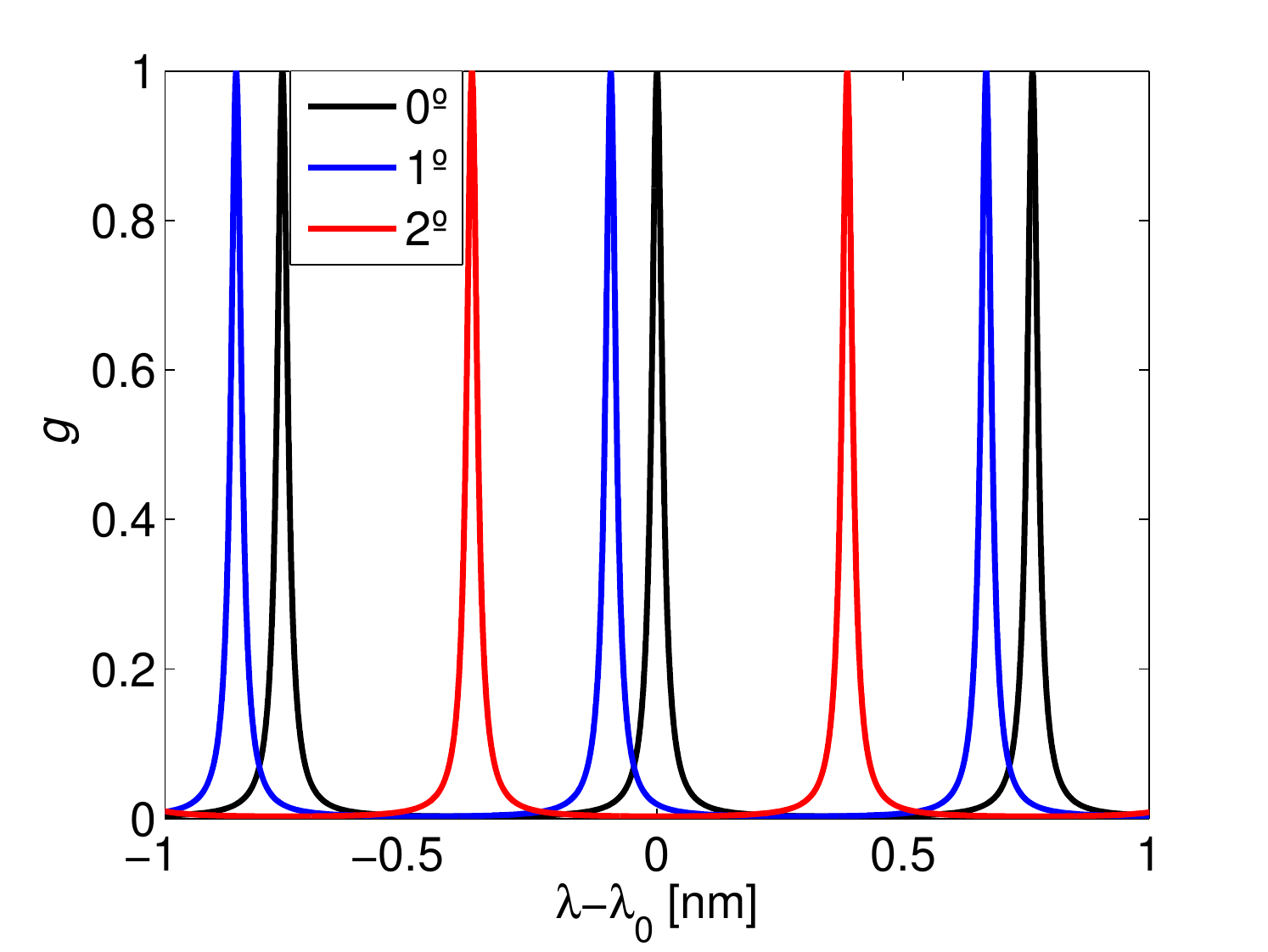}
	\end{center}
	\caption{Transmission profile of an isotropic etalon as a function of wavelength distance to $\lambda_{0}$. An incident plane wave is assumed. Black, blue, and red lines correspond to $\theta = 0^{\circ}$, $1^{\circ}$, and $2^{\circ}$, respectively.}
	\label{fig:transprof}
\end{figure}

Now one can easily realize that Eq.\ (\ref{eq:transratio}) provides a periodic function of $\delta$ whose maxima are produced when $\delta_{0} = 2 m \pi$, with $m \in \mathbb{Z}$ or, equivalently, when
\begin{equation}
\label{eq:interorder}
2n^{\prime} h \cos \theta^{\prime} = m \lambda_{0}.
\end{equation}
$m$ can be called the interferential order. A graphical representation of $g$ as a function of wavelength can be seen in Figure \ref{fig:transprof}. We have used $n\prima = 1$,\footnote{As for air at room temperature.} $h = 250$\ $\mu$m, $A = 0$, $R = 0.9$, and $\lambda_{0} = 617.234$ nm. For a given etalon with fixed refractive index and thickness, a different refraction (incidence) angle shifts the peaks of the transmission profile. Incident angles of $\theta = 0^{\circ}$ (black line), $1^{\circ}$ (blue line), and $2^{\circ}$ (red line) have been used. Simple differentiation of Eq.\ (\ref{eq:interorder}) readily shows that the peak shift is to the blue if $\theta\prima$ is increased and to the red if $\theta\prima$ is decreased.

\subsection{Properties of the transmission profile}
\label{sec:properties}

\subsubsection{Transmission peak width and order separation}
\label{sec:peakwidth}

If we call $w$ the (angular) FWHM of the peaks, it is easy to see that half the maximum is reached at 
\begin{equation}
\label{eq:deltaepsilon}
\delta_{w} = 2 m \pi \pm \frac{w}{2},
\end{equation}
or, according to Eq.\ (\ref{eq:transratio}), when 
\begin{equation}
\label{eq:efe}
\frac{1}{F} = \sin^{2} \frac{w}{4},
\end{equation}
that is, when
\begin{equation}
\label{eq:epsilon}
w = \frac{4}{\sqrt{F}},
\end{equation}
where we have assumed that $\sin (w/4) = w/4$. The FWHM in Eq.\ (\ref{eq:epsilon}) is in radians. If we want it in wavelength units, it is easy to get 
\begin{equation}
\label{eq:deltalambdaeps}
\Delta \lambda_{w} = \frac{w \lambda_{0}^{2}}{4\pi n\prima h \cos \theta\prima} = \frac{2\lambda_{0}}{\pi m \sqrt{F}},
\end{equation}
by differentiating Eq.\ (\ref{eq:phasedif}) and using Equation (\ref{eq:interorder}).

Note that $F$ in Eq.\ (\ref{eq:ef}) is an increasingly monotonic function of the reflectivity $R$. Therefore, Eq.\ (\ref{eq:deltalambdaeps}) tells us that the width of the transmission peaks basically depends on the reflectivity of the etalon. Note that $\Delta\lambda_{w}$ can slightly vary with the refraction angle (the bigger the angle, the broader the peak). This variation has small effects in solar applications as an angle of 1$^{\circ}$ represents a 2 \% modification of $\Delta \lambda_{w}$.

The {\em free spectral range} or separation between two successive peaks is equal to a shift
\begin{equation}
\label{eq:Deltadelta}
\Delta\delta_{{\rm free}} = 2 \pi.
\end{equation}
In wavelength units, analogously to Eq.\ (\ref{eq:deltalambdaeps}),  
\begin{equation}
\label{eq:deltalambdafree}
\Delta \lambda_{{\rm free}} = \frac{\lambda_{0}^{2}}{2n\prima h \cos \theta\prima} = \frac{\lambda_{0}}{m}.
\end{equation}

The free spectral range, thus, only depends on the optical thickness and on the refraction angle. The order separation without contamination of contiguous ones (a kind of cleanliness of the etalon transmission profile) is then given by the so-called {\em finesse}, 
\begin{equation}
\label{eq:finesse}
{\cal F}_{{\rm r}} \equiv \frac{\Delta \delta_{{\rm free}}}{w} = \frac{\pi \sqrt{F}}{2},
\end{equation}
which is larger when the internal reflectivity on the etalon is larger. With this definition, the FWHM of the transmission peak can be rewritten as
\begin{equation}
\label{eq:widthfinesse}
\Delta \lambda_{w} = \frac{\lambda_{0}}{m {\cal F}_{\rm r}}, 
\end{equation}
or, equivalently,
\begin{equation}
\label{eq:resolvingpower}
\frac{\lambda_0}{\Delta\lambda_{w}} = m \cal{F}_{\rm r}.
\end{equation}

The finesse is then inversely proportional to the FWHM of the transmission peaks: the larger the finesse, the thinner the peaks. The Fabry-P\'erot resolving power is directly given by the product of the interferential order and the finesse. Since the width of real etalons can change due to other factors (see Sect. \ref{sec:defectfinesse}) and the concept of finesse remains useful, ${\cal F}_{\rm r}$ in Eq. (\ref{eq:finesse}) can be called the {\em reflectivity finesse}. 
\subsubsection{Tunability of the etalon}
\label{sec:tunability}
The wavelength tuning procedure in real etalons implies a change in $n\prima$, in $h$, or in $\theta$. Equation (\ref{eq:interorder}) provides the necessary relationship between the three parameters and the wavelength of the transmission peak. If the selected tuning procedure is a tilt of the incidence angle, then one can approximately calculate that an angle
\begin{equation}
\label{eq:Deltatheta}
\Delta\theta \simeq \sqrt{\frac{\lambda_0 n\prima}{h}}
\end{equation}
is necessary to tune the etalon from one transmission peak to the next (a whole free spectral range):\footnote{This equation can be obtained by using Eq.\ (\ref{eq:interorder}) for $m$ with $\theta = 0$ and for $m+1$. For typical values of real etalons of interest in solar physics, $(1/m)^2$ turns out to be negligible (hence the approximation).} for example, with the values used for plotting Fig.\ \ref{fig:transprof}, $\Delta\theta \simeq 2\fdeg85$. 

Since Eqs.\ (\ref{eq:interorder}), (\ref{eq:deltalambdaeps}), and (\ref{eq:deltalambdafree}) depend on $\cos \theta\prima$, the transmission function is not the same across the field of view when illumination is out from normal incidence. Then, it is highly advisable to work with etalons as close as possible to normal incidence.

If we now keep fixed the incident angle, then a tuning equation can be derived from Eq.\ (\ref{eq:interorder}) by taking logarithmic derivatives:
\begin{equation}
\label{eq:tuning}
\frac{\Delta\lambda_{0} (V)}{\lambda_{0}} = \frac{\Delta n\prima (V)}{n\prima} + \frac{\Delta h(V)}{h}, 
\end{equation}
where we have assumed that the tuning agent, the driver for changing the thickness or the refractive index of the etalon is voltage. This is the case of piezoelectric or electro-optic etalons that can change either $n\prima$, $h$ or both by changing the feeding high voltage signal. 

According to \cite{ref:alvarez}, the converse piezoelectric effect in $Z$-cut crystals\footnote{Uniaxial crystals are certainly anisotropic and hence birefringent materials. We mention them here to illustrate a way of changing its (ordinary) refractive index.} can be described by the linear relationship
\begin{equation}
\label{eq:piezo}
\Delta h (V) = d_{33} V
\end{equation}
and the electro-optic change in the refractive index is given by (the unclamped Pockel's effect formula)
\begin{equation}
\label{eq:pockels}
\Delta n\prima (V) = - \frac{{n\prima}^{3} r_{13} V}{2h}.
\end{equation}

Combining Eqs.\ (\ref{eq:tuning}), (\ref{eq:piezo}), and (\ref{eq:pockels}), we get the final tuning relationship\footnote{The actual values of the $d_{33}$ and $r_{13}$ coefficients depend on the specific sample device.}
\begin{equation}
\label{eq:voltagetuning}
\Delta\lambda_{0} = \left( d_{33} - \frac{{n\prima}^{3} r_{13}}{2} \right) \frac{\lambda_{0}V}{h}.
\end{equation}

\subsubsection{Sensitivity to variations in the refractive index and etalon thickness}
\label{sec:sensitivity}

Three are the key parameters describing the etalon transmission profile, namely, the central wavelength, the peak FWHM, and the free spectral range. If the incident angle of the light beam is kept constant, according to Eqs.\ (\ref{eq:interorder}), (\ref{eq:deltalambdaeps}), (\ref{eq:deltalambdafree}), and (\ref{eq:voltagetuning}), these three parameters depend on the refractive index $n\prima$ and the thickness $h$. Impurities in the material or defects in polishing the surfaces can induce irregularities in any of them (or both) across the etalon clear aperture. These changes in the optical thickness can induce modifications in $\lambda_{0}$, $\Delta \lambda_{w}$, and $\Delta \lambda_{\rm free}$. An assessment of those possible changes is in order.

Error propagation in Eq.\ (\ref{eq:interorder}) provides
\begin{equation}
\label{eq:deltalambda0}
\frac{\delta\lambda_{0}}{\lambda_{0}} = \frac{\delta n\prima}{n\prima} + \frac{\delta h}{h}.
\end{equation}
Error propagation in Eq.\ (\ref{eq:deltalambdaeps}) provides
\begin{equation}
\label{eq:errordeltalambdaeps}
\frac{\delta(\Delta\lambda_{w})}{\Delta\lambda_{w}} = -\frac{\delta n\prima}{n\prima} - \frac{\delta h}{h}.
\end{equation}
A similar equation can be found for perturbations in the free spectral range:
\begin{equation}
\label{eq:errordeltalambdafree}
\frac{\delta(\Delta\lambda_{\rm free})}{\Delta\lambda_{\rm free}} = -\frac{\delta n\prima}{n\prima} - \frac{\delta h}{h}.
\end{equation}

Therefore, a given percent error in $h$ or $n\prima$ is transmitted directly to $\lambda_{0}$, $\Delta \lambda_{w}$, and $\Delta \lambda_{\rm free}$. Since typical thickness inhomogeneities in etalons are of the order of 1 nm, they amount a factor $4\cdot 10^{-6}$ for thicknesses of 250 $\mu$m, approximately. This is perfectly negligible for $\Delta \lambda_{w}$ and $\Delta \lambda_{\rm free}$. However, significant shifts of the order of the FWHM can be produced for the wavelength transmission peak. Perturbations in the refractive index are also much more important for the peak wavelength than for the peak width and free spectral range: a small percent or per thousand may be perfectly negligible for $\Delta \lambda_{w}$ and $\Delta \lambda_{\rm free}$ but not for $\lambda_{0}$.

In summary we can say that the expected impurities or inhomogeneities in our etalons affect less the shape of the transmission profile than the peak wavelength. See Sect.\ \ref{sec:configurations} for a discussion on these defects for the two typical optical configurations in which etalons are mounted in astronomical instruments.

\subsubsection{Transmission peak as a function of the incident angle}
\label{sec:transangle}

Let us consider a variation in $\delta$ due to a modification in the refraction angle (or the incidence angle, of course) for a given wavelength. In such a case, Eqs.\ (\ref{eq:phasedif}) and (\ref{eq:transratio}) predict a maximum of the transmission profile for normal incidence. At given wavelengths, the transmitted intensity decreases with an increasing incidence angle. This is the cause of the so-called {\em pupil apodization} that is discussed later in Sect.\ \ref{sec:telecentric}.

The monochromatic decrease in intensity is indeed induced by a shift in wavelength of the transmission peaks. Error propagation can now be written as 
\begin{equation}
\label{eq:deltalamangle}
\frac{\delta\lambda_{0}}{\lambda_{0}} = \frac{\delta \cos \theta\prima}{\cos \theta\prima} = \sqrt{1 - \frac{\sin^{2} \theta}{{n\prima}^{2}}} -1,
\end{equation}
where we have assumed shifts with respect to the peak (at $\theta = \theta\prima = 0$). If the incidence angle is small, we can write last equation in a more simple way:
\begin{equation}
\label{eq:deltalsmall}
\frac{\delta\lambda_{0}}{\lambda_{0}} \simeq -\frac{ \theta^2}{{2n\prima}^{2}}.
\end{equation}
For our sample etalon in Fig.\ \ref{fig:transprof}, a maximum incidence angle of $0\fdeg4$  translates to a maximum wavelength shift of, approximately, 15 pm, larger than the typical peak FWHM. Notice that the shift can be reduced by increasing the refraction index. For example, for Lithium Niobate, $n\simeq 2.3$ and $\delta\lambda_0\simeq2.8$ pm. Again, the effect of non-normal incident angle is negligible for the width of the transmission peaks and the free spectral range. Note that the right-hand side term of Eq.\ (\ref{eq:deltalamangle}) is $~ \, 2.4\cdot 10^{-6}$; when multiplied by $\Delta \lambda_{w} \sim 10$ pm, it gives $\delta \Delta \lambda_{w} \sim 2.4\cdot 10^{-4}$ pm. It is important to remark that, no matter the incidence angle, the right-hand sides in Eqs.\ (\ref{eq:deltalamangle}) and (\ref{eq:deltalsmall}) are always non-positive. This means that transmission peak shifts are always to the blue.

\section{Two optical configurations}
\label{sec:configurations}

Fabry-P\'erot etalons are used in solar physics in two typical optical configurations, namely, collimated and telecentric. In the first configuration the etalon is located at (or very close to) a pupil plane. In the second configuration the etalon is put very close to a focal plane. The properties and performance of the etalon are naturally different and are discussed in this Section. 

\subsection{Collimated configuration}
\label{sec:collimated}

\begin{figure}
	\begin{center}
		\includegraphics[width=0.5\textwidth]{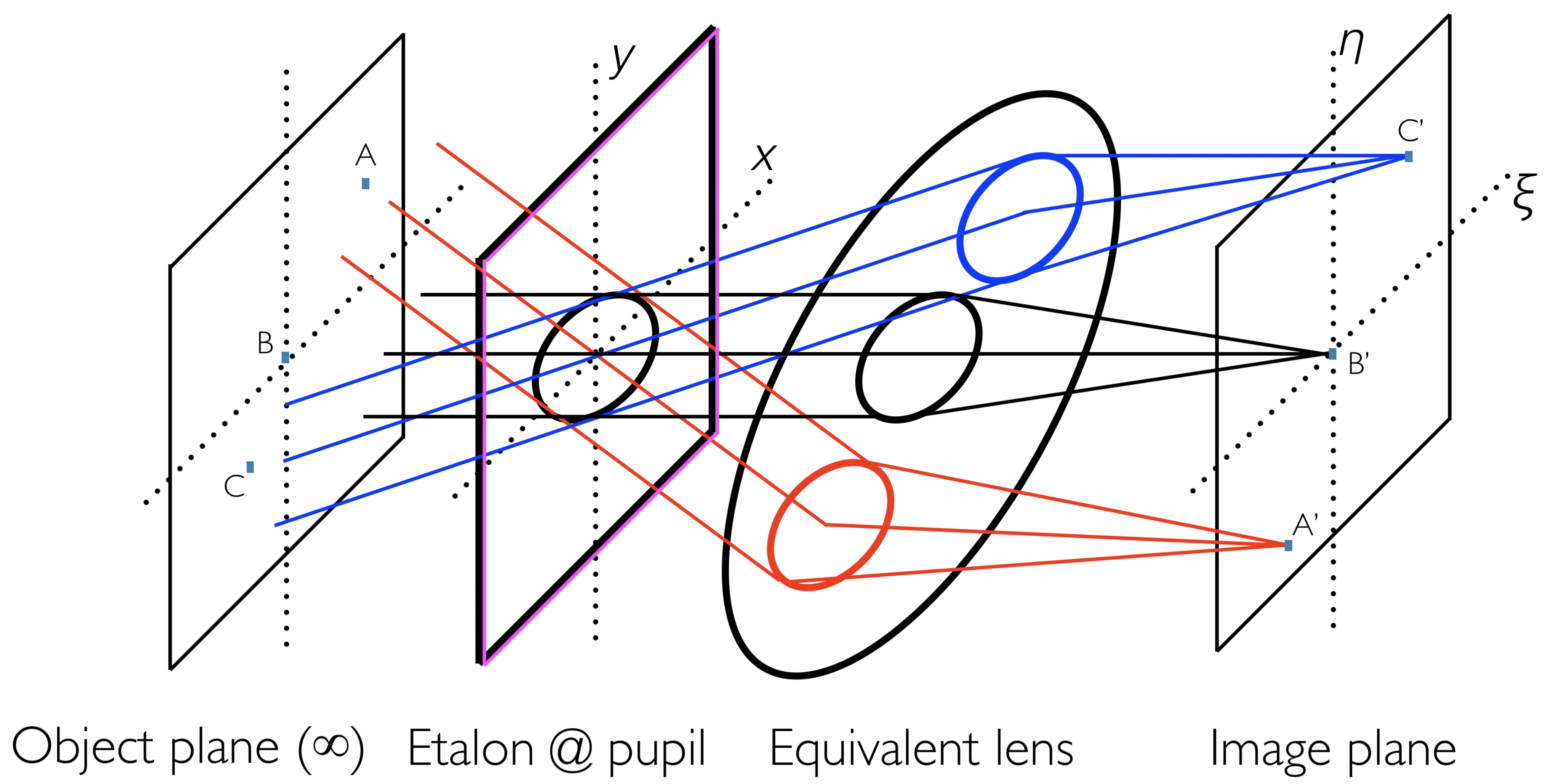}
	\end{center}
	\caption{Layout of a collimated beam etalon configuration.}
	\label{fig:collimated}
\end{figure}

Let us consider an optical configuration like the one sketched in Figure \ref{fig:collimated}. The etalon is located on a pupil plane. In such a location, the etalon is illuminated with parallel rays (plane waves) from each point in the object (assumed at infinity). The transmitted intensity at each image point is then given by Eq.\ (\ref{eq:transratio}) multiplied by the surface of the pupil. This is so because all rays added at a given image point go through the etalon with the same incidence angle. As commented on in Sect.\ \ref{sec:transangle}, we can deduce that in case of a uniform object field, images $A\prima$ and $C\prima$ will show a smaller peak intensity than $B\prima$ at a monochromatic wavelength simply because the incidence angle (hence the refraction angle) is larger. 
This is an effect that could easily be corrected for through a standard flat-field procedure. Sensors detect the flux of energy that passes through the entire transmission peak instead of the monochromatic intensity, though. As the spectral width of the profile is almost insensitive to variations in the incidence angle (Section \ref{sec:transangle}), there is no effect in the total flux of photons detected on the sensor over the field of view. What is more relevant is the wavelength shift induced by the different incidence angle. The transmission functions at points $A\prima$ and $C\prima$ are blue shifted with respect to that at $B\prima$.

The results in Sect.\ \ref{sec:transangle} account for the effects of a non-zero angle between the etalon normal and the instrument optical axis. The sensitivity of the final image to inhomogeneities of the collimated etalon can be studied with the results from Section \ref{sec:sensitivity}. Locally larger optical thicknesses imply red shifts and locally smaller optical thicknesses produce blue shifts. In this collimated configuration, the inhomogeneities are integrated and, hence, spectrally ``blurred'' on the final image. Such inhomogeneities broaden the effective transmission profile as a consequence of having different profiles shifted with respect to each other. This is discussed in the following Section \ref{sec:defectfinesse}. The consequences on the spatial point spread function of the instrument are considered in Section \ref{sec:analytical_collimated}.

\subsubsection{Effective finesse}
\label{sec:defectfinesse}

Regardless of quantitative effects, it is obvious that the highest quality etalons should be pursued. That is, we typically aim at using the smoothest, flattest, and more accurately parallel etalons. The perfect etalon does not exist, however. Defects appear in real etalons that locally change the optical path through it. Most papers and books refer to air-gapped (or other fluid) etalons and only discuss on inhomogeneities in the etalon width, $h$. Crystalline etalons, however, may also present irregularities in the refractive index, $n\prima$.\footnote{We restrict here to effects in one of the indices. Possible birefringence effects are deferred to a subsequent paper.} Since both $h$ and $n\prima$ always appear multiplied together, the relevant physical quantity is indeed the optical path $s\equiv n\prima h \cos\theta\prima$, which accounts for all possible incidence angles. The classical approach to these non-uniform etalons is to treat them as a set of individual etalons, each with a given optical thickness \citep[e.g.,][]{chabbal}. Although incoherent summation of the various etalon intensity distributions is not rigorously correct, according to \cite{ref:vaughan}, differences with the accurate coherent summation of amplitude distributions are not very large. These differences were studied by \cite{hernandez}, who showed that they are negligible for high-quality (highly reflective) etalons. Then, the common approach  \citep[e.g.,][]{ref:atherton} is to ascribe different finesses to the various plate defects under consideration and add their inverses quadratically. This was first proposed by \cite{meaburn} under the assumption that all functions involved in the degradation of the intensity profile were Gaussian. 

\begin{figure}[h]
	\begin{center}
		\includegraphics[width=0.48\textwidth]{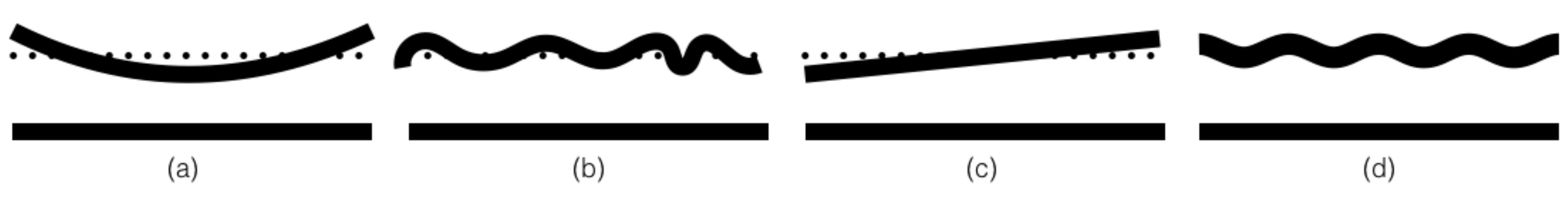}
	\end{center}
	\caption{Typical defects of Fabry-P\'erot etalons: (a) spherically shaped with a peak-to-peak excursion $\Delta s_{\rm s}$; (b) irregularities following a Gaussian distribution with a variance $\Delta s_{{\rm g}}^{2}$; (c) linear wedge with a peak-to-peak deviation $\Delta s_{p}$; and (d) sinusoidal defect of peak-to-peak amplitude $\Delta s_{a}$.}
	\label{fig:platedefects}
\end{figure}

The most commonly employed expressions for the spherical, Gaussian, and departure from parallelism finesse defects (Fig. \ref{fig:platedefects}) are, probably, those presented by \cite{chabbal}. Analytical expressions for the sinusoidal defect (Fig. \ref{fig:platedefects}d) have not been presented in the literature up to our knowledge, although this defect has already been studied by \cite{sloggett} and \cite{Hill}. Defect finesse formulas presented by \cite{chabbal} are restricted, however, to the limit when the defect distribution is very broad compared to the original transmission profile (i.e., without including irregularities). This happens either when defects are very large or when the reflectivity is high and, therefore, the original spectral profile is very narrow. The latter case is of interest as achieving high finesses is usually intended and small variations in the optical path can degrade the profile severely. We shall refer to \cite{chabbal} expressions, then, as \emph{limiting} finesses since they restrict the maximum possible finesse of the etalon. These limiting expressions are, however,  usually employed as generic ones \cite[e.g.,][]{ref:atherton,ref:gary}, i.e., as if they were valid for any magnitude of the defect.

The most complete approach to describe the etalon plate defects is, in our opinion, the one by \cite{sloggett}, who presented a general treatment applicable to any defect form or magnitude useful for etalons whose surface reflectivity is moderate to high. He  heuristically suggested that the FWHM of a defect-broadened transmission profile, $w$, is approximately given by 
\begin{equation}
w^2=w_0^2+\alpha^2\sigma_d^2,
\label{W}
\end{equation}
where $w_0$ is the width of the profile corresponding to an etalon without defects (as given by Eq.\ \ref{eq:epsilon}), $\sigma_d$ is the standard deviation of the probability density function associated to the perturbation or error in the phase $\delta$ introduced by the defects, and $\alpha$ is a coefficient that can be derived from numerical convolution of the transmission profile of a perfect etalon with the probability density function of the errors. This coefficient depends on the type and magnitude of the defect. \cite{sloggett} obtained by numerical methods that $\alpha$ converges to $2\sqrt{3}\simeq 3.46$ for all defects in the small magnitude regime ($\sigma_d/w_0<0.1$). This value of $\alpha$ agrees with the results found analytically by \cite{ref:steel}, who considered small perturbations of the incident wavefront caused by etalon defects. Note that for large defects compared to the original spectral profile ($\sigma_d>\!\!>w_0$), the width of the degraded profile is equivalent to that of the defect distribution, $w_d$, and the value of $\alpha$ coincides with the factor that relates the FWHM of the distribution with its standard deviation ($w_d=\alpha\sigma_d$). The value of $\alpha$ in this limit must be consistent with the results of \cite{chabbal}. 
 
With such a broadened profile, the reflective finesse represents no longer a spectral ``cleanliness'' of the etalon transmission profile. However, we can identify
\begin{equation}
\label{eq:defectfinesse}
{\cal F}_{d} \equiv \frac{2\pi}{\alpha \, \sigma_{d}}
\end{equation}
as a {\em defect} finesse and speak of an {\em effective} finesse given by
\begin{equation}
\label{eq:effectivefinesse}
{\cal F}_{\rm eff} \equiv \left( \frac{1}{{\cal F}_{r}^{2}} + \frac{1}{{\cal F}_{d}^{2}} \right)^{-1/2}.
\end{equation}
With this definition, we can continue using the finesse concept as an useful parameter for characterizing the etalon spectral cleanliness. Hence, using this effective finesse in Eq.\ (\ref{eq:widthfinesse}) instead of the reflectivity finesse, the actual width of the etalon transmission peak becomes
\begin{equation}
\label{eq:actualwidth}
\Delta \lambda_{w} = \frac{\lambda_{0}}{m {\cal F}_{\rm eff}}. 
\end{equation}

\cite{sloggett} pointed out that defect finesse expressions obtained through  Eq. (\ref{eq:defectfinesse})  could differ from the limiting finesses of \cite{chabbal} depending on the magnitude of the defect. He did not explicitly obtained finesse expressions for the different defects, though. We believe that they need to be presented in order to compare them with those of \cite{chabbal} and others.  We shall present here compact expressions for four examples of the defect finesse assuming defects are small ($\alpha=2\sqrt{3}$). A complete discussion on the derivation of the defect finesses is carried out in Appendix \ref{appendix}. 

Consider a dish-like defect with a spherical or parabolic shape like the one shown in Fig. \ref{fig:platedefects} (a) characterized by a peak-to-peak excursion $\Delta s_{\rm s}$ in the optical path.\footnote{\cite{sloggett} refers to the peak-to-peak excursions as $2\Delta s$ instead of $\Delta s$.} The defect finesse can be shown to be given by, 
\begin{equation}
{\cal F}_{d_{\rm s}}=\frac{\lambda}{2\Delta s_{\rm s}}.
\label{eq:spherical}
\end{equation}

If we now focus on Fig.\ \ref{fig:platedefects} (b), we have a micro-rough surface with deviations from $s$ that follow a normalized Gaussian distribution with variance $\Delta s_{\rm g}^{2}$. In this case, the defect finesse is
\begin{equation}
{\cal F}_{d_{\rm g}}=\frac{\lambda}{4\sqrt{3} \Delta s_{\rm g}}\simeq\frac{\lambda}{6.9\Delta s_{\rm g}}.
\label{eq:gaussian}
\end{equation}

Third, if departure from parallelism is linear as in Fig.\ \ref{fig:platedefects} (c), with a peak-to-peak excursion of $\Delta s_{\rm p}$, then the defect finesse can be written as
\begin{equation}
{\cal F}_{d_{\rm p}}=\frac{\lambda}{\sqrt{3}\Delta s_{\rm p}}\simeq\frac{\lambda}{1.7\Delta s_{\rm p}}.
\label{eq:departure}
\end{equation}

Consider finally an etalon with an optical path roughness given by a sinusoid of amplitude $\Delta s_{\rm a}$ and zero offset. The corresponding defect finesse is

\begin{equation}
{\cal F}_{d_{\rm a}}=\frac{\lambda}{\sqrt{6}\Delta s_{\rm a}}\simeq\frac{\lambda}{2.5\Delta s_{\rm a}}.
\label{eq:sinusoidal}
\end{equation}

Should the defects of a given etalon be described by the superposition of two or more of these distributions, it is naturally understood that its inverse square finesse would result from summing up the square inverse finesses of each distribution.

\begin{figure} [t]
	\begin{center}
		\includegraphics[width=0.48\textwidth]{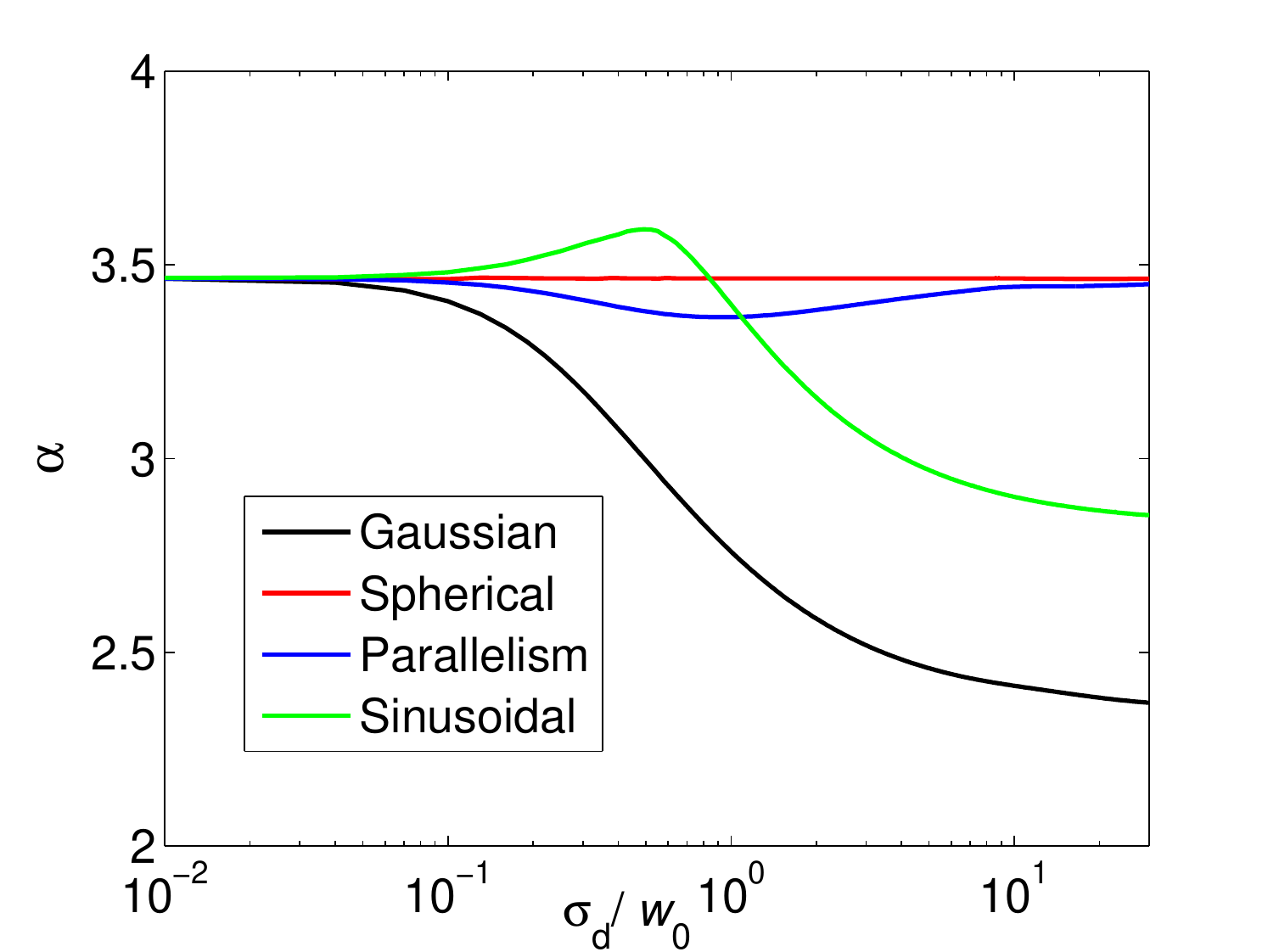}
		\caption{Value of the coefficient $\alpha$ for the Gaussian (black), spherical (red), parallelism (blue), and sinusoidal (green) defects against the standard deviation of probability density function associated to each defect normalized by the width of the profile of an etalon without defects, $\sigma_d/w_0$.}
		\label{Slogget}
	\end{center}
\end{figure}

As indicated before, for $\sigma_d>\!\!>w_0$, the value of $\alpha$ should give rise to consistent finesse expressions compared to the ones found by \cite{chabbal}. Furthermore, these are, in principle, different from Eqs. (\ref{eq:spherical}-\ref{eq:sinusoidal}). Figure \ref{Slogget} shows the value of $\alpha$ in the range $0.01<\sigma_d/w_0<30$  obtained after numerical convolution of the four defect distributions here considered (Appendix \ref{appendix}) with the transmission profile $g$ of an etalon with reflectivity $R=0.95$ and unity transmission factor.\footnote{Note that \cite{sloggett} presented the value of $\alpha$ up to $5\sigma_d/w_0$ in his paper employing a Lorentzian function as transmission profile instead of $g$.  We believe that this upper limit of $\sigma_d/w_0$ is insufficient to evaluate the tendency of $\alpha$ in the regime $\sigma_d>\!\!>w_0$. For that reason, we have extended by a factor of six.} We observe that $\alpha$ tends in all cases to $2\sqrt{3}\simeq 3.46$ for $\sigma_d/w_0<0.1$, as already shown by \cite{sloggett}. In the limit $\sigma_d>\!\!>w_0$, $\alpha$ tends to $2\sqrt{3}$ for the  spherical and parallelism distributions, to  $2\sqrt{2\ln2}\simeq 2.35$ for the Gaussian distribution and to $2\sqrt{2}\simeq 2.83$ for the sinusoidal one. The limiting finesse ${\cal F}_{d}^{\rm lim}$ coincides then with Eqs. (\ref{eq:spherical}) and (\ref{eq:departure}) for the spherical and the parallelism defects as the \emph{limiting} value of $\alpha$ coincides with that of the small defect regime. On the contrary, for the Gaussian and sinusoidal distribution the limiting finesse formulas differ from Eqs. (\ref{eq:gaussian}) and (\ref{eq:sinusoidal}). Their expressions are given by

\begin{equation}
{\cal F}_{d_{\rm g}}^{\rm lim} \simeq \frac{\lambda}{4.7\Delta s_g},
\end{equation}
and

\begin{equation}
{\cal F}_{d_{\rm a}}^{\rm lim} \simeq \frac{\lambda}{2\Delta s_a}.
\end{equation}

The limiting value of $\alpha$ coincides in each case with the factor that relates the FWHM with the standard deviation of the defect distributions (Appendix \ref{appendix}) and agrees with the limiting finesse expression of of \cite{chabbal}, as expected.

The defect finesse expressions here presented have been restricted only to two limits: ``small'' and ``large'' defects. In general, Eq. (\ref{eq:defectfinesse}) must be applied with the value of $\alpha$ that corresponds to the magnitude of the particular defect (Figure \ref{Slogget}).

\subsubsection{Transmission profile widths across the image}
\label{sec:widthacross}

A further effect can produce a differential broadening of the transmission peaks of the etalon across the focal plane in a collimated configuration. Since any point in the final image is formed with rays that went through the etalon at higher incidence angle for greater radial distances from  image center, the transmission peak broadening is dependent on such a radial distance.

Differentiating Eq.\ (\ref{eq:deltalambdaeps}), one easily gets that the relative variation in the FWHM of the peak is
\begin{equation}
\label{eq:widthratio}
\frac{\delta\Delta\lambda_{w}}{\Delta\lambda_{w}} = \tan \theta\prima \, \delta\theta'.
\end{equation}
With a typical value less than $0\fdeg5$ for the maximum incidence angle in solar telescopes, the ratio is $8\cdot 10^{-5}$. Therefore, we can safely disregard this effect for our very slow instruments.

\subsubsection{Deviations from perfect collimation}
\label{sec:imperfect_collimated}
Deviations from perfect collimation can be viewed as illuminating the etalon with a spherical wavefront of a finite numerical aperture. The consequence would be a broadening and a displacement of the profiles with respect to that of parallel illumination. Of course, aberrations can also be present in the incident wavefront, but these will not be considered here. Following \cite{sloggett} method, the broadening of the transmission profile due to the angular spread illumination of each point of the etalon can be dealt with an {\em aperture finesse} (Appendix \ref{appendix}) given by 

\begin{equation}
\label{eq:aperturefinesse}
{\cal F}_{d_{\rm f}} \equiv \frac{2\pi }{m \Omega}\frac{\nprimac}{n^2} = \frac{2 }{m \tan^{2} \theta_{\rm m}}\frac{\nprimac}{n^2},
\end{equation}
where $\Omega$ stands for the solid angle of the cone of rays traversing the etalon, and $\theta_{\rm m}$ is the maximum incidence angle in the cone.  This expression is compatible with that presented by \cite{ref:atherton}, except for the factor $\nprimac/n^2$. We think the disagreement is due to the fact that \cite{ref:atherton} considered an air-gapped etalon in their derivation and not the general (crystalline) case.

\subsection{Telecentric configuration}
\label{sec:telecentric} 

\begin{figure}
	\begin{center}
		\includegraphics[width=0.5\textwidth]{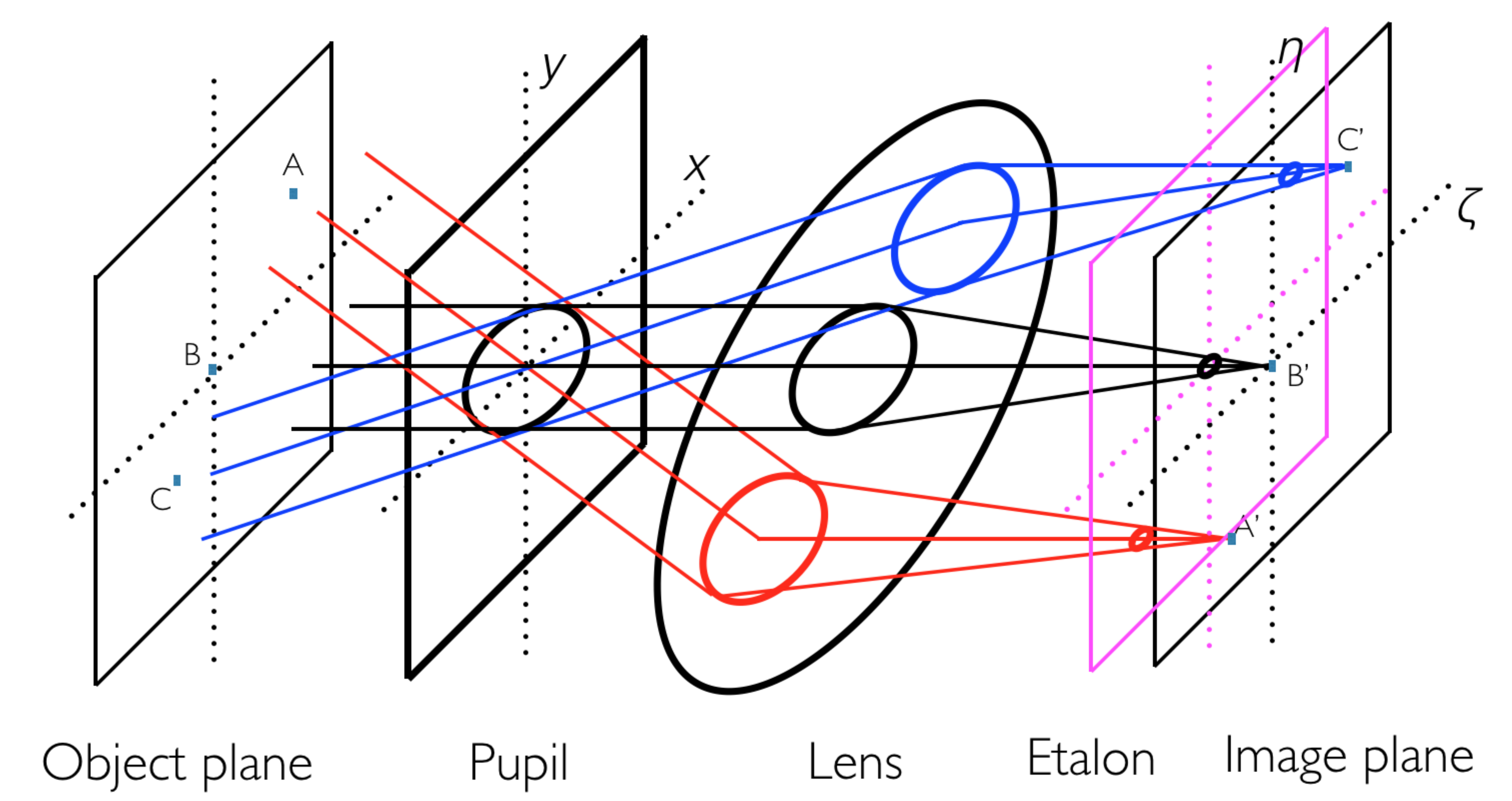}
	\end{center}
	\caption{Layout of a telecentric beam etalon configuration where the entrance pupil is located at the focus of the lens (and the exit pupil is therefore at infinity). In this case, points $A\prima$ and $C\prima$ receive the same cone of rays as for point $B\prima$.}
	\label{fig:etalontelecen}
\end{figure}

To keep the same passband across the FOV, an alternative configuration can be used. In a (image-space) telecentric configuration (Fig.\ \ref{fig:etalontelecen}) the etalon is located (almost) at the focal plane and the exit pupil at infinity (or, equivalently, the entrance pupil is at the front focal point of the system).

Each point of the etalon sees the same cone of rays coming from the pupil. Unlike the collimated case, all three $A\prima$, $B\prima$, and $C\prima$ points are evenly illuminated if the object field is flat and no wavelength shifts in the transmission peaks are expected from one point to another. The transmitted intensity is not $g$ any longer, though. Since these rays are coherent because they come from the same object point, addition of intensities does not provide a solution and we should deal with electric field amplitudes.
\subsubsection{Transmission profile}
The vector electric field of the ray transmitted by the etalon in Fig. \ref{fig:Etalontransref} is given by 
\begin{equation}
\label{eq:vectorelectric}
{\bf E}^{\rm (t)} = \frac{T{\rm e}^{{\rm i}\delta/2}}{1 - R \, {\rm e}^{{\rm i} \delta}} \, {\bf E}^{\rm (i)},
\end{equation}
where (t) and (i) refer again to the transmitted and incident quantities and $R$ and $T$ are given by Eqs.\ (\ref{eq:reflectivity}) and (\ref{eq:rplust}). This expression differs from that presented in most text books \citep[e.g.,][]{ref:born} by the general phase factor ${\rm e}^{{\rm i}\delta/2}$, which is irrelevant in their discussion. However, it is, at first, important in our current analysis as it depends on the incidence angle.\footnote{Neither \cite{vonderluhe} nor \cite{ref:Scharmer} take this phase factor into account.} The origin of the global phase is discussed in Appendix \ref{appendix2}.

\begin{figure}
	\begin{center}
		\includegraphics[width=0.48\textwidth]{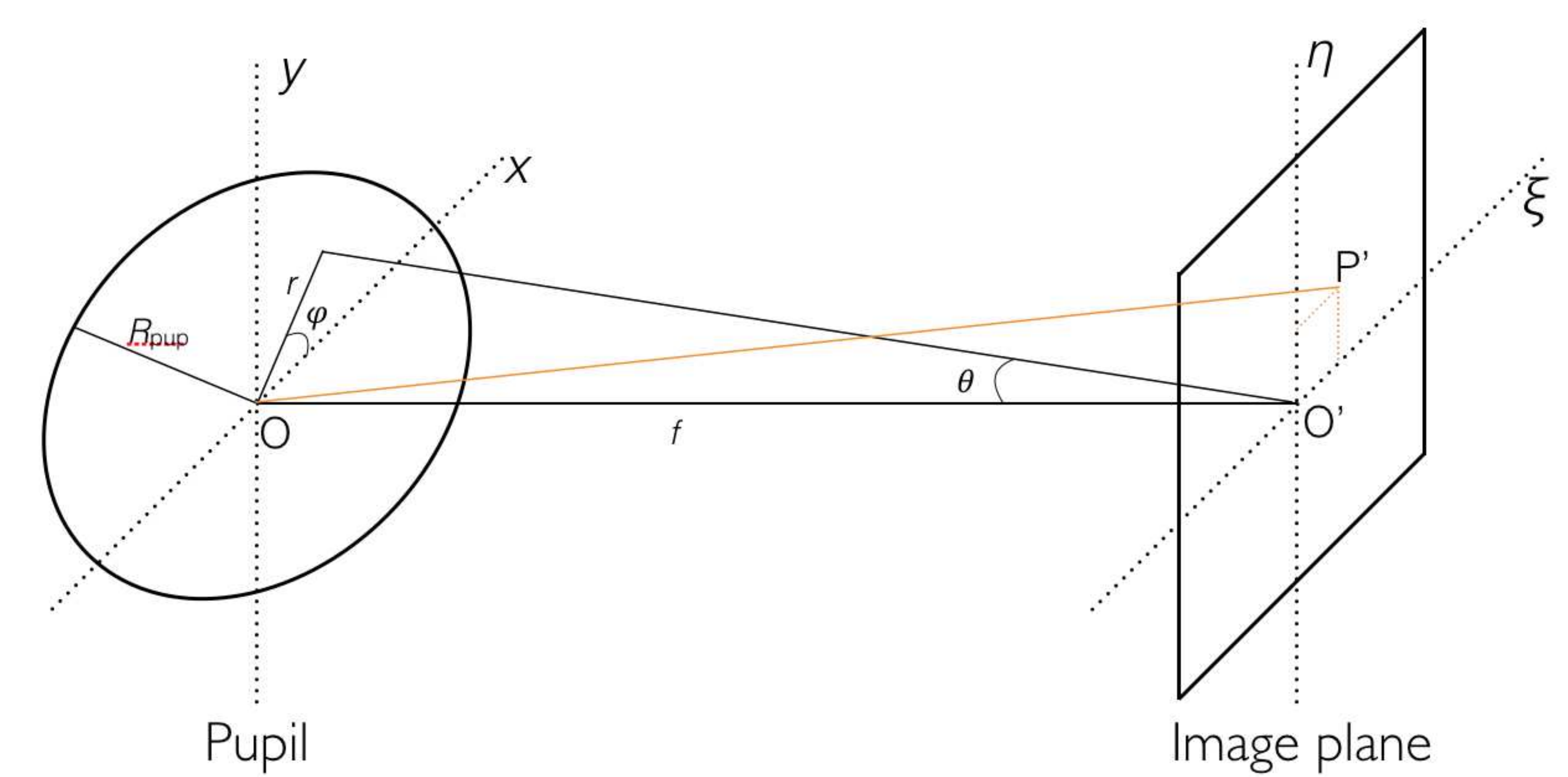}
	\end{center}
	\caption{Rays coming from the pupil to the image plane. Indeed they go from the lens in Fig.\ \ref{fig:etalontelecen} to the etalon.}
	\label{fig:pupilayetalon}
\end{figure}

With some simple algebra, Eq.\ (\ref{eq:vectorelectric}) can be cast as 
\begin{equation}
\label{eq:vectorelectric2}
{\bf E}^{\rm (t)} = \frac{\sqrt{\tau}}{1 - R}  \, \frac{ {\rm e}^{{\rm i} \delta/2} - R \, {\rm e}^{{\rm -i} \delta/2}}{1 + F \sin^{2} (\delta/2)} \, {\bf E}^{\rm (i)},
\end{equation}
where $\tau$ is defined in Eq.\ (\ref{eq:tau}) and $F$ in Equation (\ref{eq:ef}). Consider now the geometry sketched in Figure\ \ref{fig:pupilayetalon}. For a general optical system, the electric field at any point $P\prima = (\xi,\eta)$ is given by the sum of all electric fields across the pupil surface:
\begin{equation}
\label{eq:fourier}
\begin{gathered}
\tilde{{\bf E}}^{\rm (t)} (\xi, \eta) = \frac{1}{\pi R_{\rm pup}^2} \, \int\!\!\!\!\int_{\rm pupil} {\bf E}^{\rm (t)} (x,y) \, {\rm e}^{-{\rm i} k(\alpha x + \beta y)} \, {\rm d}x \, {\rm d}y, 
\end{gathered}
\end{equation}
where $\alpha \equiv \xi/f$ and $\beta \equiv \eta/f$ are the cosine directors of $OP\prima$ (notice that we restrict ourselves to small angles).\footnote{We have normalized by the pupil surface in order to obtain quantities that can later be compared with the results for the collimated configuration.} Therefore, the electric field in the image plane is proportional to the Fourier transform of that in the pupil plane. For our discussion about the telecentric configuration we should  concentrate in the electric field at point $O\prima$: all the points in the focal plane receive the same cone of light.\footnote{It is interesting to remark that there is not a general convention on the (arbitrary) positiveness or negativeness of $\delta$ and, consequently, on the sign of the exponent of the direct Fourier transform. Other authors, such as \cite{Hecht} use an opposite sign to the one used here.}

The axial symmetry of Fig.\ \ref{fig:pupilayetalon} indicates that the pupil electric field only depends on $r$ and we can thus write 
\begin{equation}
\label{eq:fieldatzero}
\tilde{{\bf E}}^{\rm (t)} (0,0) = \frac{2}{R_{\rm pup}^{2}} \, \int_{0}^{R_{\rm pup}} r\, {\bf E}^{\rm (t)} (r) \,{\rm d}r.
\end{equation} 
All points in the pupil at a distance $r$ from its center have an associated incidence angle $\theta$ to the etalon. Therefore, each monochromatic ray out of the optical axis contributes less and less energy (Sect.\ \ref{sec:transangle}) as $\theta$ increases. The bigger the distance to the pupil center, the smaller the energy. Hence the pupil is seen from the etalon as if it were not evenly illuminated. This is the so-called {\em pupil apodization} effect, first discovered by \cite{ref:beckers} and later discussed and elaborated by \cite{vonderluhe} and \cite{ref:Scharmer}. Moreover, those rays coming from the external parts of the pupil have their corresponding transmission peaks shifted to the blue (Sect. \ref{sec:transangle}) with respect to the central ray. Therefore, the integration of all rays should translate into a blue shifted and a broadened transmission peak with the subsequent loss of spectral resolution as compared to the collimated case.

The average ratio between the transmitted and incident intensities in the telecentric configuration is then given by
\begin{equation}
\label{eq:transratio2}
\tilde{g} = \frac{\tilde{{\bf E}}^{\rm (t)} {\tilde{\bf E}}^{{\rm (t)} \ast}} {{\bf E}^{\rm (i)} {\bf E}^{{\rm (i)} \ast}}, 
\end{equation}
where the asterisk indicates the complex conjugate. 
\begin{figure}
	\begin{center}
		\includegraphics[width=0.48\textwidth]{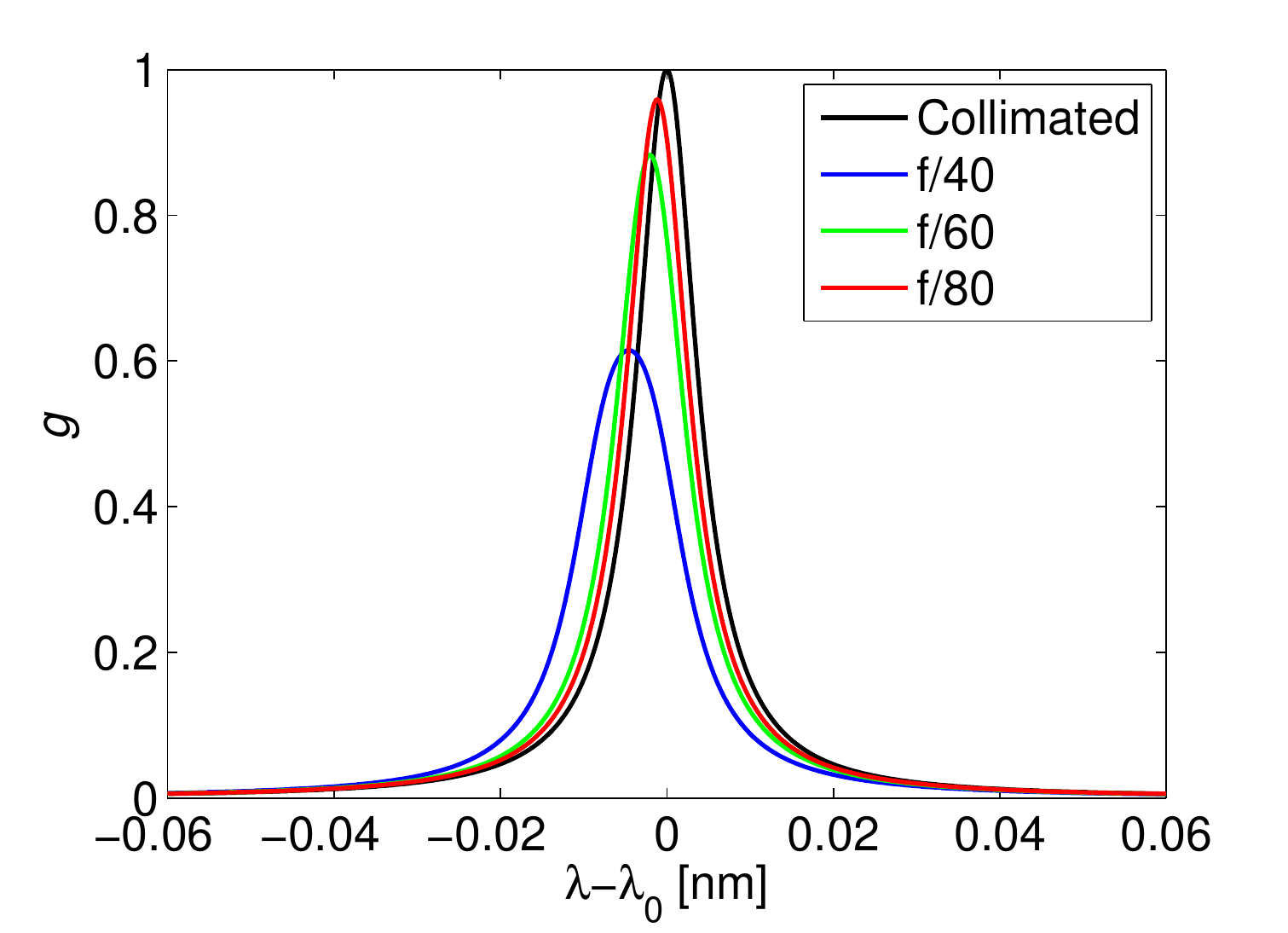}
	\end{center}
	\caption{Transmission profiles as functions of the wavelength distance to $617.28$ nm. A collimated configuration of the etalon is represented in black line. A telecentric configuration with $f/40$ (blue), $f/60$ (green) and $f/80$ (red) is also shown.}
	\label{fig:transcoltel}
\end{figure}
Figure \ref{fig:transcoltel} shows a plot of the average transmission peak in a telecentric configuration with $f/40$, $f/60$ and $f/80$.
As a reference, the same etalon but in a collimated configuration is used.
A refractive index of $n' = 2.3$ has been used along with $\lambda_{0} = 617.28\,$nm, $h = 250\, \mu$m, $A = 0$, and $R=0.92$.
We will employ these parameters, corresponding to a commercial etalon, throughout the rest of this work. Table 1 gives the remaining key parameters for evaluating $g$ and $\tilde{g}$ after Eqs.\ (\ref{eq:transratio}) and (\ref{eq:transratio2}), respectively. As expected, the transmission profiles reduce their peak intensity and broaden when changing from the collimated configuration to telecentric configuration. The transmission profiles are also shifted bluewards with respect to the reference wavelength. These effects are more prominent for smaller f-numbers due to the increasing of the aperture of the incident cone of rays.

\subsubsection{Effects on the effective finesse and on the peak wavelength}
\label{sec:finiteaperture}

To circumvent the tedious (rigorous) calculation of Eq. (\ref{eq:fieldatzero}) after having substituted the electric field of Eq. (\ref{eq:vectorelectric2}) into it, we can use the aperture finesse of Sect. \ref{sec:imperfect_collimated} as an approximate measure of the transmission profile broadening. One should only include ${\cal F}_{d_{\rm f}}$ in the effective finesse expression of Equation (\ref{eq:effectivefinesse}). To assess the validity of such an approximation, we re-write ${\cal F}_{d_{\rm f}}$ in terms of the image space f-number, $f\#$, as

\begin{equation}
{\cal F}_{d_{\rm f}} =\frac{8 (f\#)^{2}}{m}\frac{\nprimac}{n^2},
\label{F_fnum}
\end{equation}
which gives $\cal{F}_{\rm eff} =$ 36.5, 34.2, 26.2 for $f/80, f/60, f/40$, respectively. These values are to be compared with the exact ones given in Table 1. As expected, the larger the $f\#$, the better the approximation.

We have seen that another consequence of receiving a cone of rays instead of a collimated beam is a blue shift of the spectral profile (Sect.\ \ref{sec:telecentric}). From the average change of phase compared to the collimated case, it can be shown (Appendix \ref{appendix}) that the spectral shift of the profile, $\Delta\lambda_0$,  depends on both the refraction index and the f-number through 
\begin{equation}
\Delta\lambda_0\simeq -\frac{\lambda_0}{16(f\#)^2}\frac{n^2}{\nprimac}.
\label{shift}
\end{equation}

That is, for larger f-numbers and refraction indices, the spectral shift decreases. For a collimated beam, $f\#\rightarrow\infty$, we have $\Delta\lambda_0\rightarrow 0$ and ${\cal F}_{\rm a}\rightarrow \infty$, as expected. Using this equation, the expected blue shifts are about $-4.55$ pm, $-2.02$ pm, and $-1.14$ for $f/40$, $f/60$, and $f/80$. These values fit extraordinarily well with those presented in Table \ref{tab:results1}.\footnote{\cite{ref:title} found the same analytical expression for the blue shift of the spectral profile. The derivation he followed is not rigorous though, since it is based on an analytical expression for the transmitted profile obtained by averaging Eq. (\ref{eq:transratio}) over the cone of rays instead of adding electric field amplitudes.}

\begin{table}
	\begin{center}
		\caption{Etalon parameters in four configurations, namely, collimated, and telecentric with f/80, f/60, and f/40.}
		\label{tab:results1}
		\vspace{0.2truecm}
		\begin{tabular}{lcrrr}
			\hline 
			{ } \\
			Parameters & Collimated & $f/80$ & $f/60$  & $f/40$\\
			{ } \\
			\hline
			\hline
			$\tau$ & 1 & 0.96 & 0.88 & 0.60\\
			$\Delta\lambda_{0}$ (pm) & 0 & -1.13 & -2.02 & -4.55\\
			$\Delta\lambda_{w}$ (pm) & 8.80 & 9.18 & 9.97 & 13.9\\
			$\Delta\lambda_{\rm free}$ (nm) & 0.33 & 0.33 & 0.33 & 0.33\\
			${\cal F}_{\rm eff}$ & 37.7 & 36.1 & 33.2 & 23.8\\
			\hline
		\end{tabular}
	\end{center}
\end{table}

\subsubsection{Plate-defect-induced effects}
\label{sec:deftel}
An assessment on how sensitive the final image is to etalon inhomogeneities and to a non-zero angle between the instrument optical axis and the etalon optical axis is as easy as in Sections\ \ref{sec:sensitivity} and \ref{sec:transangle}. Equations (\ref{eq:errordeltalambdaeps}) and (\ref{eq:errordeltalambdafree}) are the same for all rays in the incoming cone of light because they are independent of the incident angle in a telecentric configuration. Then, the average transmitted intensity in Eq.\ (\ref{eq:transratio2}) will suffer exactly the same effect across the image, namely, that defects or errors in the optical thickness are only important for the wavelength tuning of the transmission peak. Modifications in the FWHM of peaks and the free spectral range can be neglected. Equation (\ref{eq:deltalamangle}) is also valid for all the rays in the cone. Hence, we should only take care of changes in the peak wavelength.

Since the defects of the etalon are directly mapped to the image in this telecentric configuration, the wavelength shifts have a direct influence in the derived LOS velocities with the instrument. To correct at first order for these LOS velocity shifts one can measure them while taking flat-field exposures: if we determine the line position for every pixel with a flat illumination, we should only subtract the so-derived velocities from those evaluated independently. However, it is important to remark that the induced artificial LOS velocities cannot be corrected completely in telecentric mounts unless the PSF is fully characterized both spatially and spectrally. This is probably one of the most important disadvantages of this configuration.

\section{The PSF in the two configurations}\label{sec:psf}
Let us now study the spectral and spatial PSF of the telecentric configuration compared to the collimated case.

Equations (\ref{eq:vectorelectric2}) and (\ref{eq:fourier}) are fully general for both configurations since they hold for monochromatic plane waves impinging the etalon. The electric field on the image plane is the Fourier transform of that illuminating the pupil. The difference between the two systems is whether ${\bf E}^{\rm (t)}$ and the phase difference $\delta$ are  constant across the pupil or not. That is, they are  independent of the spatial coordinates $(x,y)$ of the pupil plane in the collimated configuration whereas they are not in the telecentric configuration: ${\bf E}^{\rm (t)}$ and $\delta$ do depend on $x$ and $y$.

\subsection{PSF in collimated configuration}\label{sec:analytical_collimated}

\begin{figure}
	\begin{center}
		\includegraphics[width=1.01\columnwidth]{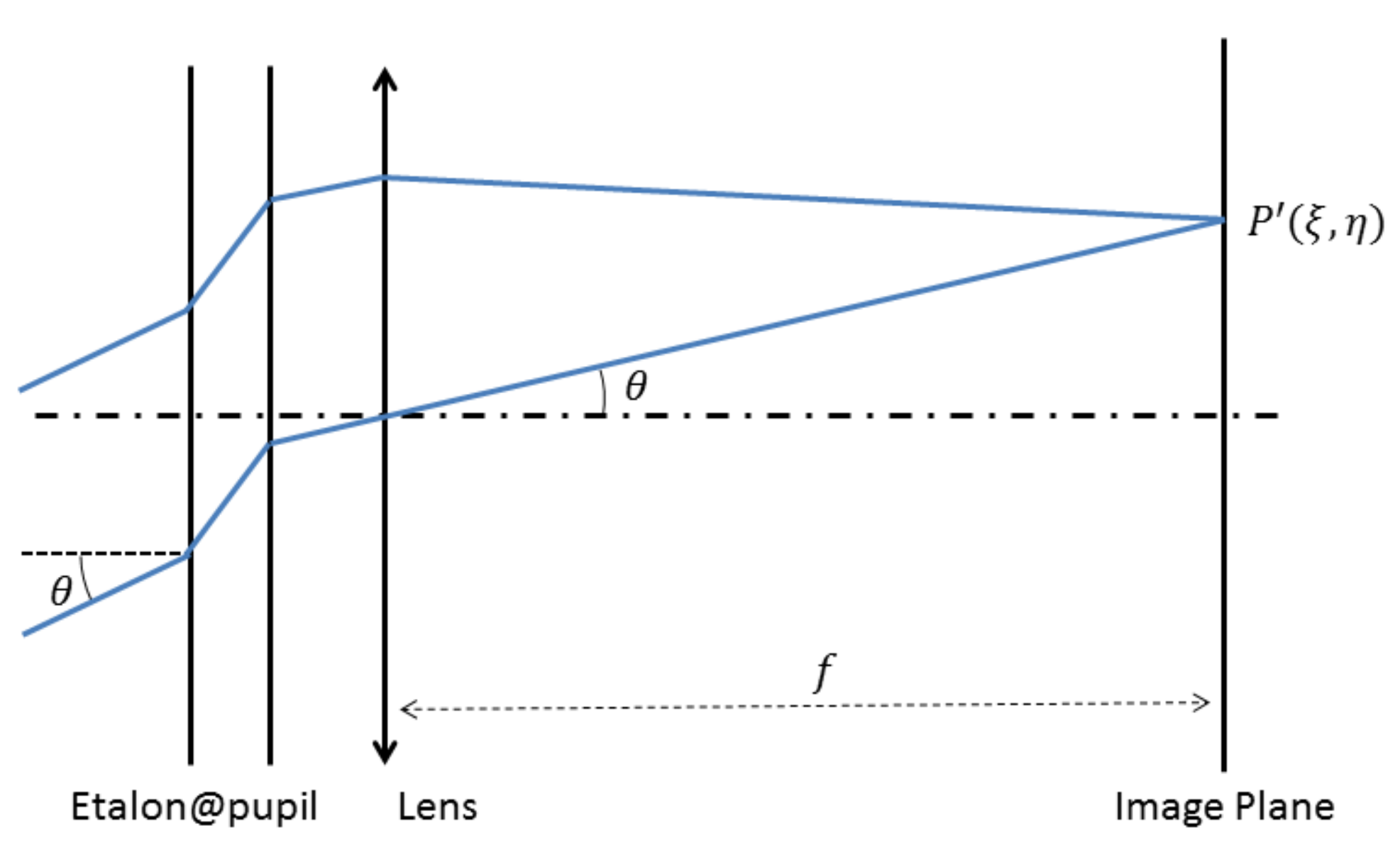}
	\end{center}
	\caption{2D layout of a collimated beam etalon configuration for two rays that impact on the etalon with an angle $\theta$.}
	\label{colimada}
\end{figure}

Figure \ref{colimada} displays a 2D layout of a collimated etalon configuration where two rays of incidence angle $\theta$ reach the etalon and, later, the image plane. Since the etalon is placed on the pupil, all rays striking on it with an angle $\theta$ will be projected on the same point $P'(\xi,\eta)$ of the image plane no matter their incidence positions at the etalon. A relationship between $P'$ and the incidence angle $\theta$ can easily be found if we assume the stop is placed at the object nodal plane of the system (in a single lens paraxial system, this means that the stop is placed at the lens and the central ray is not deviated). If $f$ stands for the focal length,

\begin{equation}
\cos\theta=\frac{f}{\sqrt{\xi^2+\eta^2+f^2}}.
\label{cos_coll}
\end{equation}

The phase difference $\delta$ at $P\prima$ can be written then as
\begin{equation}
\delta (\xi, \eta) = \frac{4\pi h}{\lambda} \, \sqrt{n'^2 -n^2 + \frac{n^2f^2}{\xi^2+\eta^2+f^2}}.
\label{delta_coll}
\end{equation}

It is important to remark that $\delta$ does not depend on the pupil plane coordinates $(x,y)$.  Therefore, for a perfect etalon with no defects, Eq. (\ref{eq:fourier}) simply turns into

\begin{equation}
\Etilde(\xi,\eta)=\frac{1}{\pi R_{\rm pup}^2} \Et_0(\xi,\eta)\int\!\!\!\!\int_{\rm pupil}{\rm e}^{-ik(\alpha x+\beta y)}\text{d}x\text{d}y,
\label{Et_coll}
\end{equation}
where $\Et_0 (\xi,\eta)$ is the electric field transmitted by the etalon that approaches $P\prima$. It should be noticed that Eq. (\ref{Et_coll}) is proportional to the Fraunhofer integral of a circular aperture \citep{Hecht}. Hence,\footnote{Here we use $z_0$ as a parameter, which is denoted by the semicolon in front of it.}

\begin{equation}
\tilde{\bf E}^{\rm (t)} (z;z_0) = \tilde{\bf E}_0^{\rm (t)} (z_0) \frac{2J_1 (z-z_0)}{z-z_0},
\label{Etilde}
\end{equation}
where $J_1(z)$ is the first order Bessel function and the variable $z$ is given by

\begin{equation}
z=\frac{2\pi}{\lambda}R_{\rm pup}\frac{\sqrt{\xi^2+\eta^2}}{f}.
\end{equation}

Unlike in the case of a clear circular aperture, space invariance has been lost with the collimated etalon and the response of the system depends on the position across the image. Thus, the point spread function cannot be interpreted as a regular PSF since it varies from point to point. The instrument does not respond with the convolution of the object intensity distribution with the PSF. Rather, one has to multiply the object surface brightness with the {\em local} PSF and integrate. Such a local PSF can be expressed as
\begin{equation}
\PSF(z;z_0) = g(z_0) \, \left[\frac{2J_1(z-z_0)}{z-z_0} \right]^2,
\end{equation}
where $g(z_0)$ is given by Eq. (\ref{eq:transratio}) with the dependence on $z_0$ given through Equation (\ref{delta_coll}). Then, the monochromatic, local PSF turns out to be  the same as the PSF produced by a circular aperture except for a transmission factor. 
 This result enables to interpret the response of the etalon as that of a clear circular aperture (hence, through convolution with ${\cal S}_0 \equiv [2J_1(z)/z]^2$) but multiplied with the local transmission profile value. In other words, we have an \emph{apodization} of the image. This implies that an etalon without defects in collimated configuration only affects the image quality by reducing the {\em monochromatic} intensity. As soon as we go radially out from the optical axis, $g(z_0)$ is shifted in wavelength (see Section 4.1) and, hence, it is reduced compared to the transmission factor ($\tau$) at the given wavelength. Therefore, the most significant consequence we can expect of image apodization is a radial decrease of the monochromatic $S/N$ of the observations since the largest noise source is typically photon noise, which is proportional to the square root of the signal. Since $g$ is a monotonically decreasing function of $z$, longer focal lengths can be beneficial for given etalons at the expenses of either reducing the FOV or increasing the size of the detector. 

So far we have discussed the monochromatic behavior of the etalons. Our instruments always integrate a finite passband per each wavelength sample; thus, the polychromatic response has to be addressed. This is done in Section \ref{sec:quasi1}.

For a real etalon with defects, Eq. (\ref{Et_coll}) is no longer valid.
Either $h$, $n'$, or both depend on the pupil plane coordinates since the defects are located at specific points $(x,y)$. This dependence must be incorporated into Eq. (\ref{delta_coll}) and the PSF should be evaluated numerically.
 An approximation of the real PSF can be obtained through the convolution of $\PSF$ with a defect density distribution much in the same way as we do in order to get the results of Section \ref{sec:defectfinesse}. 
As ${\cal S}_0$ does not depend on $\delta$, such a convolution can only affect $g(z_0)$. We can then safely expect that the net effect of inhomogeneities are mostly seen in the spectral transmission, but not in the spatial shape of the PSF. \footnote{Attention must be paid if the Strehl's ratio is used for evaluating the wavefront degradation in etalons since small variations in the optical path can lead to large variations in the transmission in $g(z_o)$. Thus, a decay in the monochromatic Strehl's ratio may come from a decay in the monochromatic transmission and not from  degradation of the PSF. In our opinion, the Strehl's ratio should be employed only with the quasi-monochromatic PSF (Sec. \ref{sec:quasi1}). In any case, the PSFs normalization factors need to be chosen taking into account that the energy enclosed by the degraded and unaberrated PSFs must be the same.} 

On the other hand, the (unavoidable) presence of micro-roughness errors in the reflecting surfaces should translate into an increase of the energy contained in the wings of the PSF, as they are high-frequency errors. This undesired excess of energy in the lobes of the PSF is commonly referred as \emph{stray light} and its consequence is a loss of contrast. In spectropolarimetry, stray light is a particularly delicate issue, though, because it represents a contamination of the magnetic signal at a given feature by the signal originated in other structures located all around the feature.

Consider now an imperfectly collimated input beam. The phase shift depends in this case on the pupil coordinates as the incidence angle changes across the etalon. The net effect is essentially the same as locating the etalon in an imperfect telecentric configuration (Section \ref{sec:imperfect}). This is obvious as we only care about the irradiance distribution across the detector and, thus, the integrals that must be performed are the same as in the telecentric case except for a an irrelevant scale factor that accounts for the projection of the pupil on the etalon. The only difference is that etalon defects are still averaged out over the illuminated area, whereas in the telecentric mount defects are directly mapped into the detector.
\subsection{PSF in telecentric configuration}
\label{sec:telecentricPSF}
\begin{figure}
	\begin{center}
		\includegraphics[width=1.05\columnwidth]{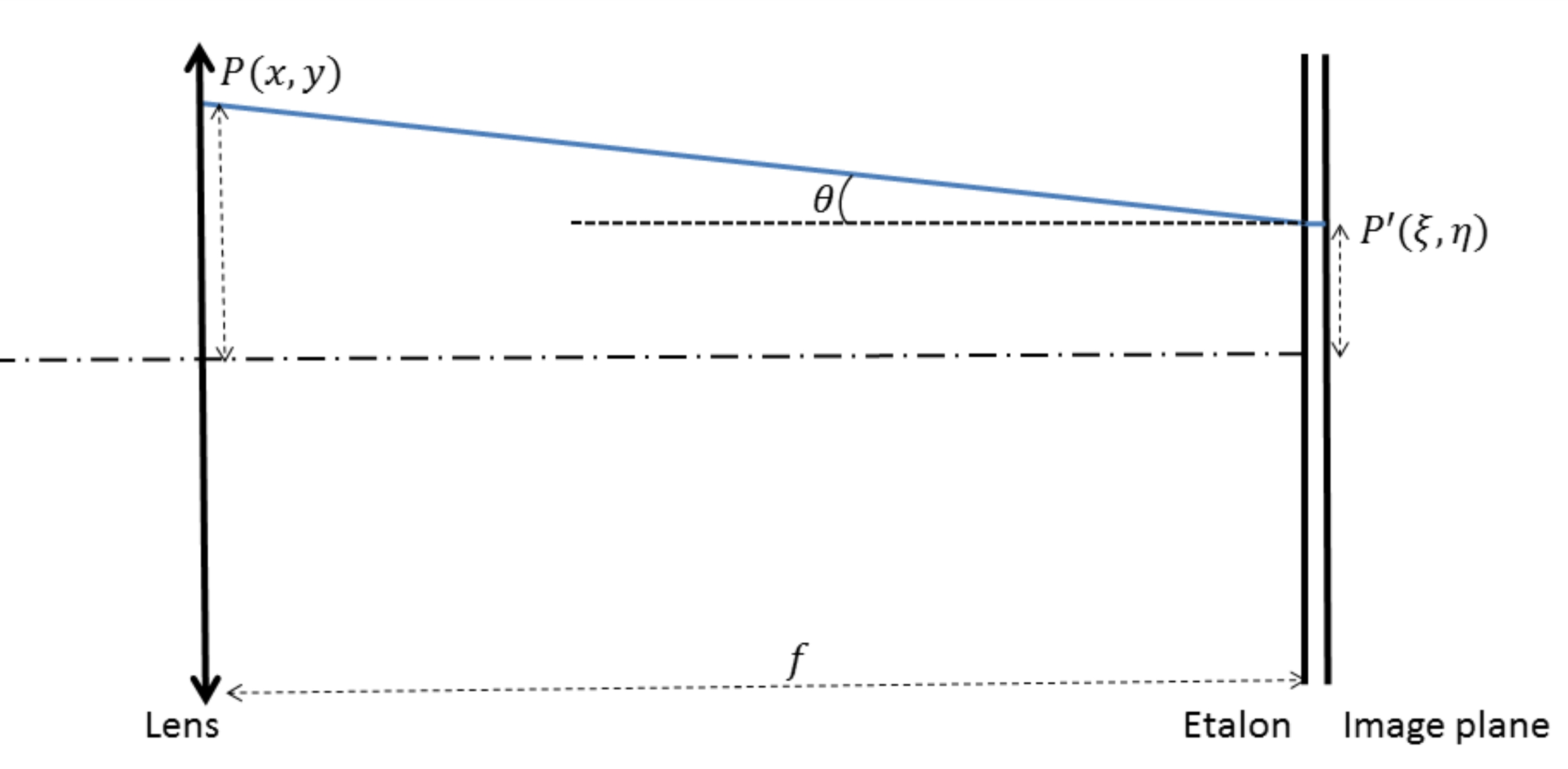}
	\end{center}
	\caption{2D Layout of a telecentric beam etalon configuration for a ray that comes from the pupil at $P(x,y)$ and is projected to the etalon at $P'(\xi,\eta)$.}
	\label{telecentric}
\end{figure}

In the telecentric configuration, any point $P\prima(\xi,\eta)$ of the etalon sees a cone of rays, each coming from different parts of the pupil. 
Therefore, the phase shift $\delta$ now depends as well on the pupil plane coordinates. From Fig. \ref{telecentric},

\begin{equation}
\cos \theta=\frac{f}{\sqrt{(x-\xi)^2+(y-\eta)^2+f^2}}.
\end{equation}

Hence, the explicit dependence on both the pupil and image plane coordinates is

\begin{equation}
\begin{gathered}
\delta (x,y,\xi,\eta)=\\
=\frac{4\pi h}{\lambda} \sqrt{\nprimac-n^2+ \frac{n^2f^2}{(x-\xi)^2+(y-\eta)^2+f^2}}.
\end{gathered}
\label{delta_tel}
\end{equation}

Likewise in the collimated etalon with defects case, ${\bf E}^{\rm (t)}$ does depend here on the pupil plane coordinates and cannot be taken out from the integral in Equation (\ref{Et_coll}). The PSF must be calculated numerically.

\begin{figure}
	\begin{center}
		\includegraphics[width=0.48\textwidth]{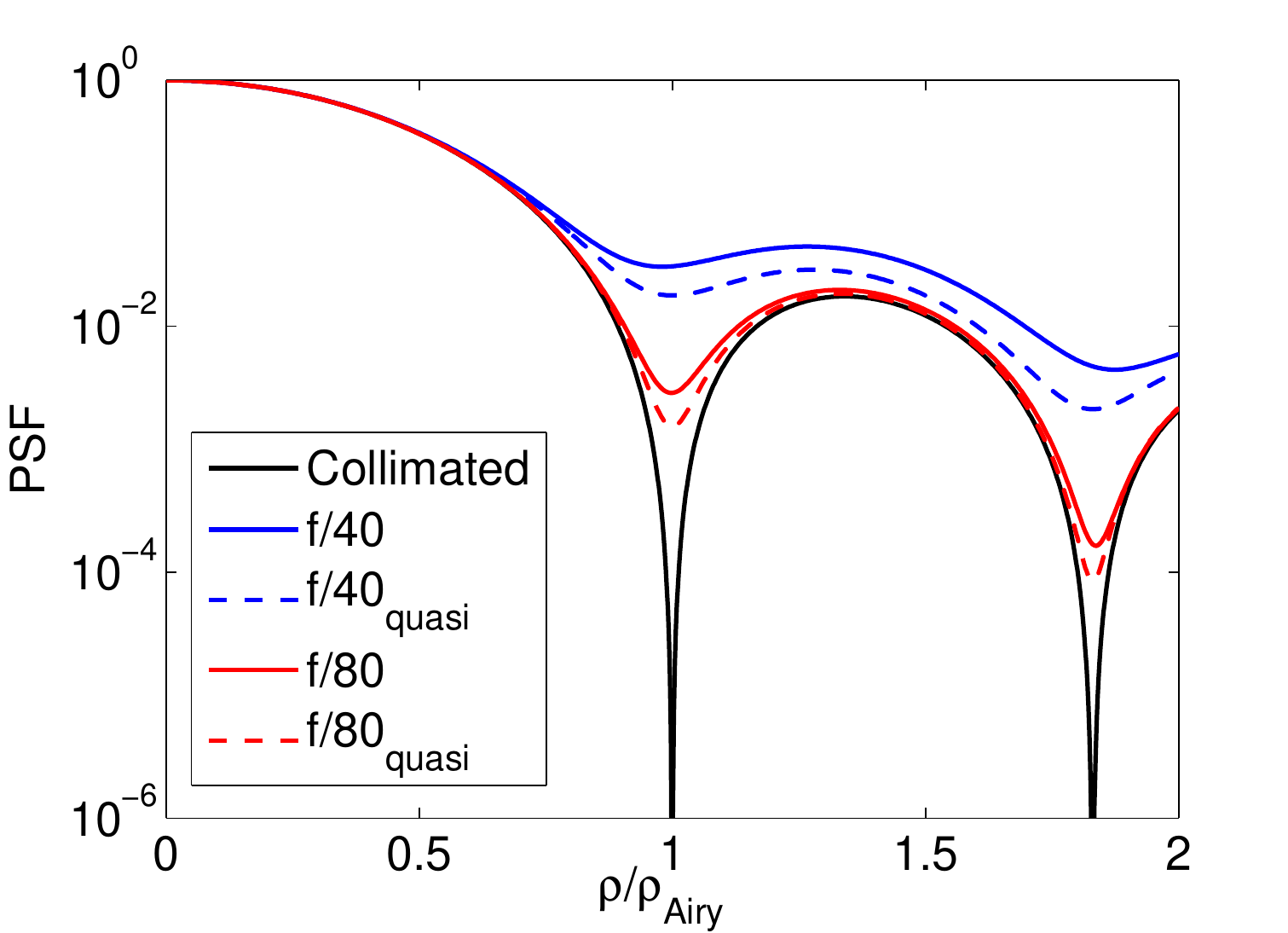}
	\end{center}	
	\caption{Normalized PSFs calculated in the telecentric configuration at f/40 and f/80 (blue and red line respectively) and in the collimated configuration (black) line for normal illumination of the pupil. The quasi-monochromatic PSFs of both f-numbers have also been represented (blue and red dashed lines respectively).}
	\label{PSFlog_centered}
\end{figure}

Figure \ref{PSFlog_centered} shows the monochromatic PSFs as functions of the radial distance from the optical axis, $\rho \equiv (\xi^2+\eta^2)^{1/2}$, normalized by the Airy disk radius of a clear, circular aperture, $\rho_{\rm Airy} = 1.22 f \lambda_0(2R_{\rm pup})^{-1}$. Solid lines represent the monochromatic PSFs as evaluated at their respective peak wavelengths, $\lambda_t \equiv \lambda_0 + \Delta \lambda_0$, where $\Delta \lambda_0$ is given in Table 1. Dashed lines represent the quasi-monochromatic PSFs after integrating the finite etalon passband (see Section \ref{sec:quasi1}). Blue, and red correspond to the $f/40$ and $f/80$ telecentric cases. For the sake of comparison, the PSFs are normalized to their maximum transmissions, which are also given in Table \ref{tab:results1}.

Differences between both collimated and telecentric configurations become more evident from the vicinity of the first minimum of the Airy pattern and are more prominent, as expected, for the shorter $f\#$ beams.

\begin{figure} 
	\begin{center}
		\includegraphics[width=\columnwidth]{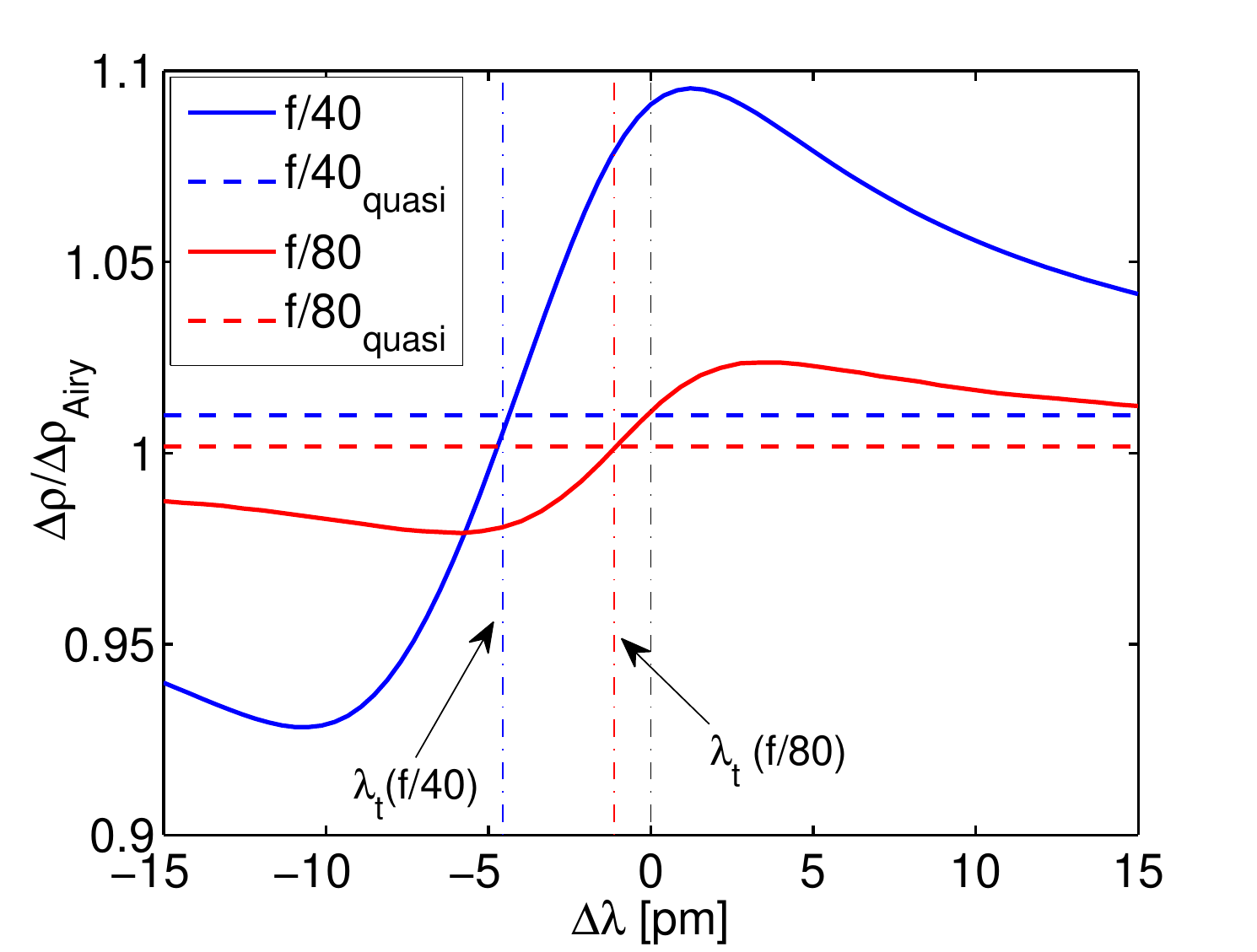}
	\end{center}
	\caption{FWHM of the PSF in a perfect telecentric configuration normalized to the Airy disk FWHM as a function of the wavelength shift . F-numbers $f/40$ (blue line) and  $f/80$ (red line) have been employed. The FWHM of the  quasi-monochromatic PSFs for $f/40$ and $f/80$ (blue and red dashed lines respectively) have also been included. Vertical, dashed-dotted lines mark the position of the maximum transmission wavelengths. In black, that of the collimated configuration.}
	\label{PSF_spectral}
\end{figure} 

\begin{figure} 
	\begin{center}
		\includegraphics[width=0.49\textwidth]{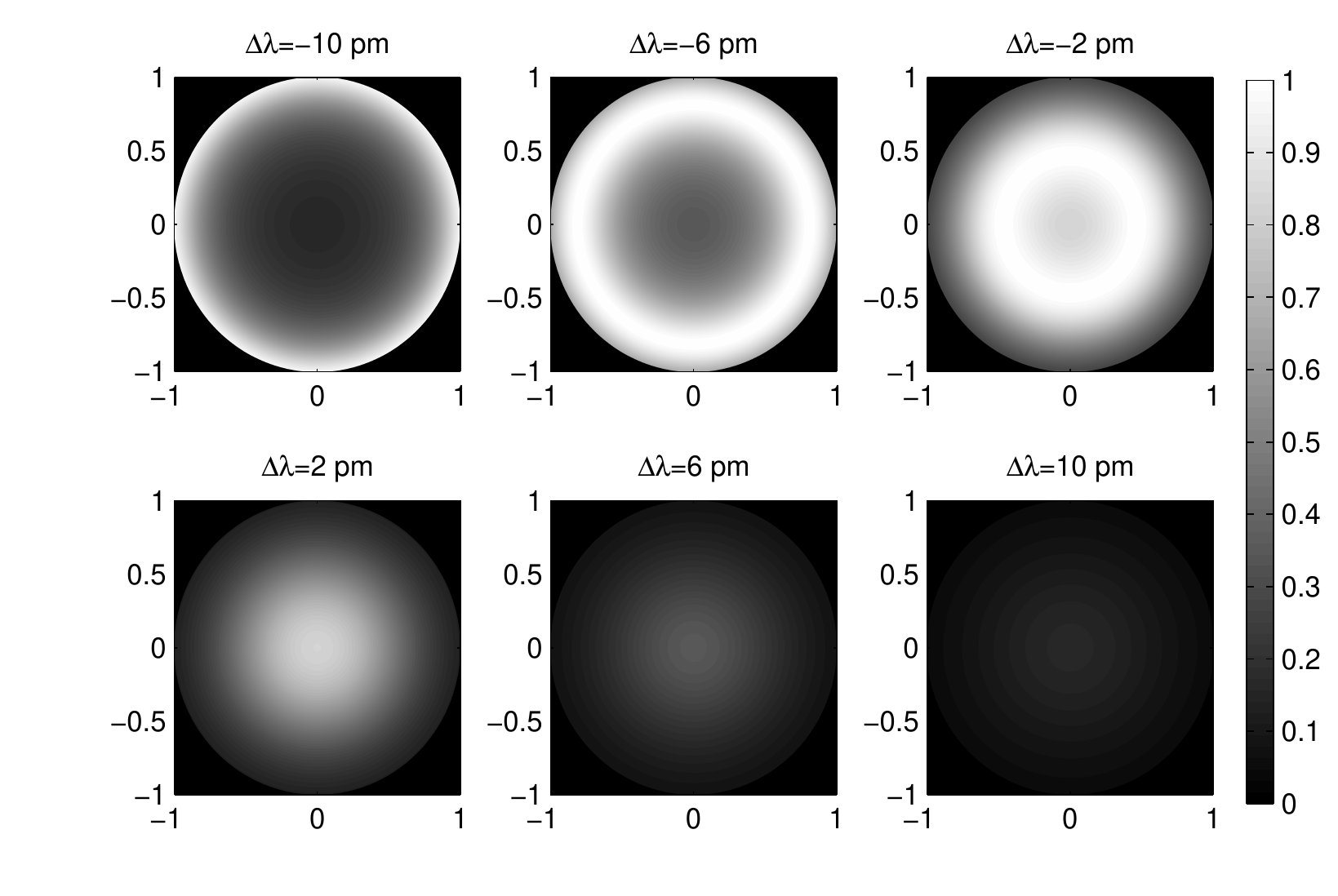}
	\end{center}
	\caption{Pupil apodization in a telecentric mount illuminated with a f/40 beam for different shifts with respect to $\lambda_0$. From the upper left to the lower right: $\Delta\lambda=-10, -6, -2, 2, 6$ and $10$ pm.}
	\label{apodization_wavelength}
\end{figure}

Following Sect. 4.2, one could expect that the telecentric PSF gets broadened as compared to the collimated case, whose width coincides with that for a clear, circular aperture: $\Delta \rho_{\rm Airy} = 1.029 f\lambda_0(2R_{\rm pup})^{-1}$. This is actually true only at certain wavelengths.
 Figure \ref{PSF_spectral} shows in solid lines the FWHM of the monochromatic PSF, $\Delta\rho$, normalized to $\Delta\rho_{\rm Airy}$, against the wavelength shift from $\lambda_0$ for both $f/40$ (blue) and $f/80$ (red). 
We can observe that the PSF broadening is a wavelength dependent effect, as evaluated for the first time by \cite{ref:beckers}. 
 The PSF narrows towards the blue with respect to the FWHM at $\lambda_t$. The opposite is the case for red wavelength displacements. The reason for this is that pupil apodization (and phase errors) is a wavelength-dependent effect (Fig. \ref{apodization_wavelength}). 
Towards the red of $\lambda_t$, the center of the pupil is brighter than the edges. The effect is very similar to a Gaussian apodization of the pupil, which translates to a broadening of the central disk of the PSF. The ``effective'' size of the pupil decreases and also reduces the energy in the secondary rings \citep{ref:mahajan}; 
 towards the blue of $\lambda_t$, a central obscuration appears and the brightness shifts with annular shape towards the edges. The practical effect of obscuring an optical system is to decrease the central disk of the PSF at the expenses of expanding the wings of the PSF \citep{ref:mahajan}, thus contributing to stray light effects. This argumentation is consistent with the results found by \cite{vonderluhe}. 

The maximum and minimum FWHM of the PSFs differ in less than a $10\%$ and $3\%$ from $\Delta\rho_{\rm Airy}$ for the f/40 and f/80 beams, respectively. Also notice that the separation between the minimum and the maximum is of the order of the FWHM of the spectral profile (Table \ref{tab:results1}). For larger shifts, the pupil tends to be evenly illuminated and the PSF of a diffraction-limited system with the same pupil size is recovered. As remarked by \cite{ref:beckers},  the wavelength dependence of the FWHM introduces artificial velocity signals in solar images with velocity structure. An evaluation of this effect in real instrumentation is presented in the third part of this series of papers, where we show that errors in the magnetic field can also appear.

In an ideal telecentric configuration, where all chief rays across the FOV are parallel to the optical axis, each point of the etalon receives the same cone of rays. Thus, all results obtained for normal illumination are also valid for any direction of the incident illumination of the pupil. 
\begin{figure}
	\begin{center}
		\includegraphics[width=\columnwidth]{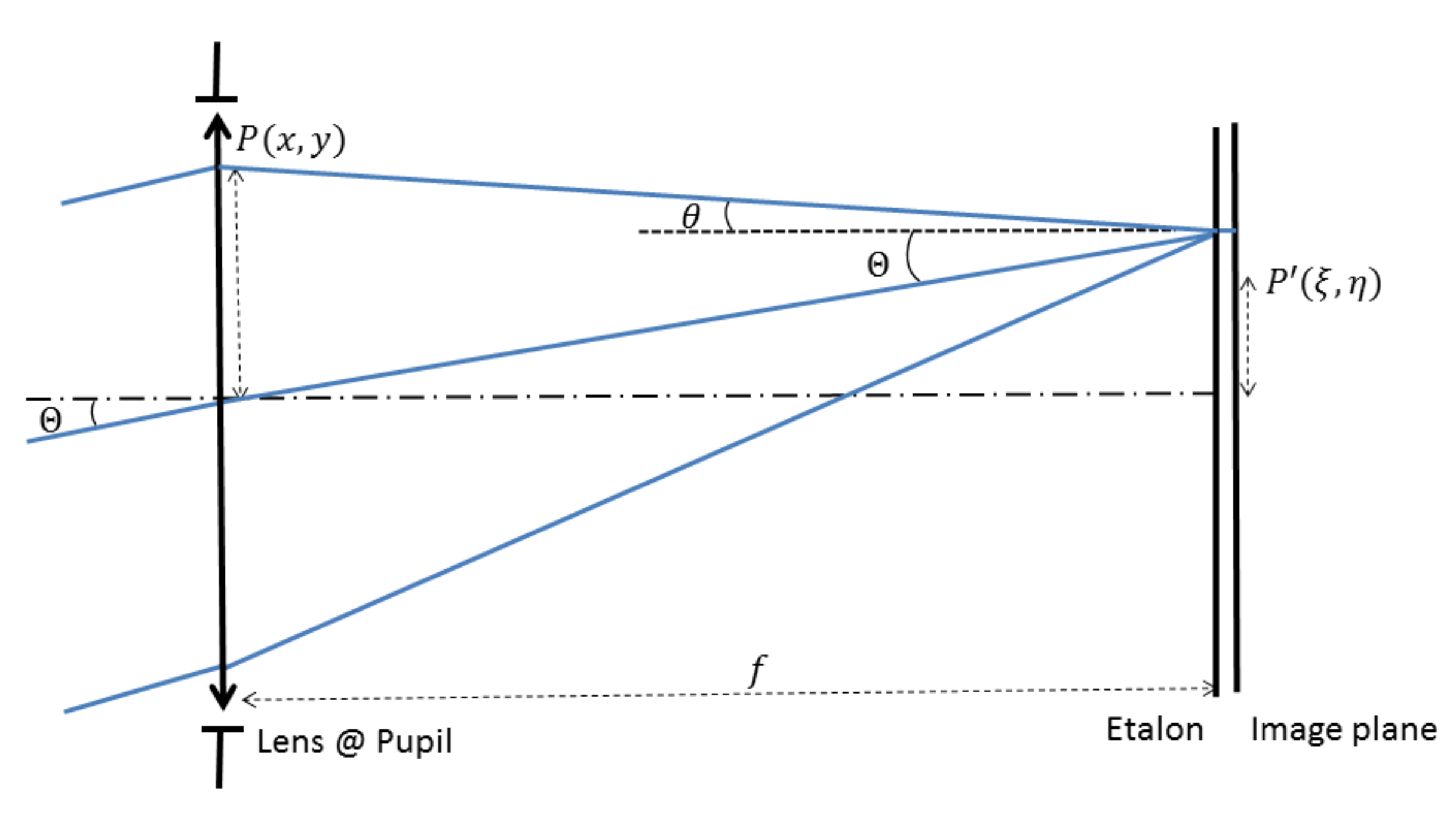}
	\end{center}
	\caption{2D Layout of a non-telecentric beam configuration (the lens and the pupil are located at the same position) for a collimated beam that illuminates the pupil with an incident angle $\Theta$. The chief ray does not deviate, whereas the rest of rays reach the etalon with different angles $\theta$.}
	\label{telecentric2}
\end{figure}

\subsection{Quasi-monochromatic PSF}
\label{sec:quasi1}
Real observations are polychromatic. We should be interested therefore in the quasi-monochromatic response of the system. Typically, in front of the quasi-monochromatic Fabry-P\'erot etalon, instruments have an order-sorting pre-filter. Let $T(\lambda)$ be the transmission profile of the pre-filter (typically a window-shape function). If $O(\xi,\eta; \lambda)$ denotes the monochromatic brightness distribution of the object, then the image quasi-monochromatic intensity distribution centered at $\lambda_0$ can be expressed as

\begin{multline}
\label{eq:intencol}
I(\xi, \eta; \lambda_0) \\= \int_{-\infty}^{+\infty}\!\!\!\!\!\!\!\!\!\!\! T(\lambda) \iint\!\!\! O(\xi, \eta; \lambda - \lambda_0) \cdot {\cal S} (\xi-\xi_0, \eta-\eta_0,\lambda-\lambda_0) \,{\rm d}\lambda\,{\rm d}\xi_0\,{\rm d}\eta_0,
\end{multline}
in the collimated configuration and 
\begin{equation}
\label{eq:intentel}
I(\xi, \eta; \lambda_0) = \int_{-\infty}^{+\infty}\!\!\!\!\!\!\!\!\!\!\! T(\lambda)\, \left[ O(\xi, \eta; \lambda - \lambda_0) \ast {\cal S} (\xi, \eta; \lambda - \lambda_0) \right] {\rm d}\lambda,
\end{equation}
in the telecentric configuration, where the symbol $\ast$ stands for convolution. Convolution in Eq.~(\ref{eq:intentel}) is carried out in the space domain. Therefore, only if the object brightness distribution is independent of wavelength \citep{vonderluhe}, as in the case of the continuum, then $O$ can go out from the integral and write
\begin{equation}
\label{eq:intentel2}
I(\xi, \eta; \lambda_0) = O(\xi, \eta; ) \ast \int_{-\infty}^{+\infty}\!\!\!\! T(\lambda)\, {\cal S} (\xi, \eta; \lambda - \lambda_0) \, {\rm d}\lambda,
\end{equation}
in the telecentric configuration. Hence, the right-hand side of the convolution can be identified as a {\em quasi-monochromatic} PSF, ${\cal S}_{\rm quasi}$, which coincides with the integral in wavelength of the monochromatic $S$ multiplied by $T(\lambda)$. 

 The quasi-monochromatic PSF is strictly  valid only for the continuum wavelengths, though. Within the spectral lines, the spatial and spectral properties of light can be highly correlated and, thus, space invariance no longer holds. The response of the instrument, then depends on the object itself. However, one can reasonably expect that the integration in wavelength somehow reduces the purely monochromatic effects in the final images at other wavelength samples. This can only be checked numerically.\footnote{We refer the reader to the third paper of this series for a quantitative evaluation of this phenomenon.}

Along with the monochromatic PSFs, Fig. (\ref{PSFlog_centered}) shows ${\cal S}_{\rm quasi}$ for the two telecentric cases in dashed lines. (The collimated ${\cal S}_{\rm quasi}$ exactly coincides with the monochromatic one after normalization.) You can see that the quasi-monochromatic PSF performance is better than that of the monochromatic one, as best witnessed close to the minima. The reason for this is that the the position of the monochromatic PSF minima are very sensitive to wavelength variations in the vicinities of $\lambda_0$. The net effect is an improvement of the PSF when averaging spectrally the monochromatic PSFs \citep{vonderluhe}. 

Figure \ref{PSF_spectral} also shows the quasi-monochromatic cases in dashed lines. The quasi-monochromatic PSF widths are larger than in the collimated configuration, although it can be seen that the effect of integrating the monochromatic PSFs virtually balances out their spectral variations.

\section{Deviations from perfect telecentrism}
\label{sec:imperfect}
\begin{figure}
	\begin{center}
		\includegraphics[width=0.5\textwidth]{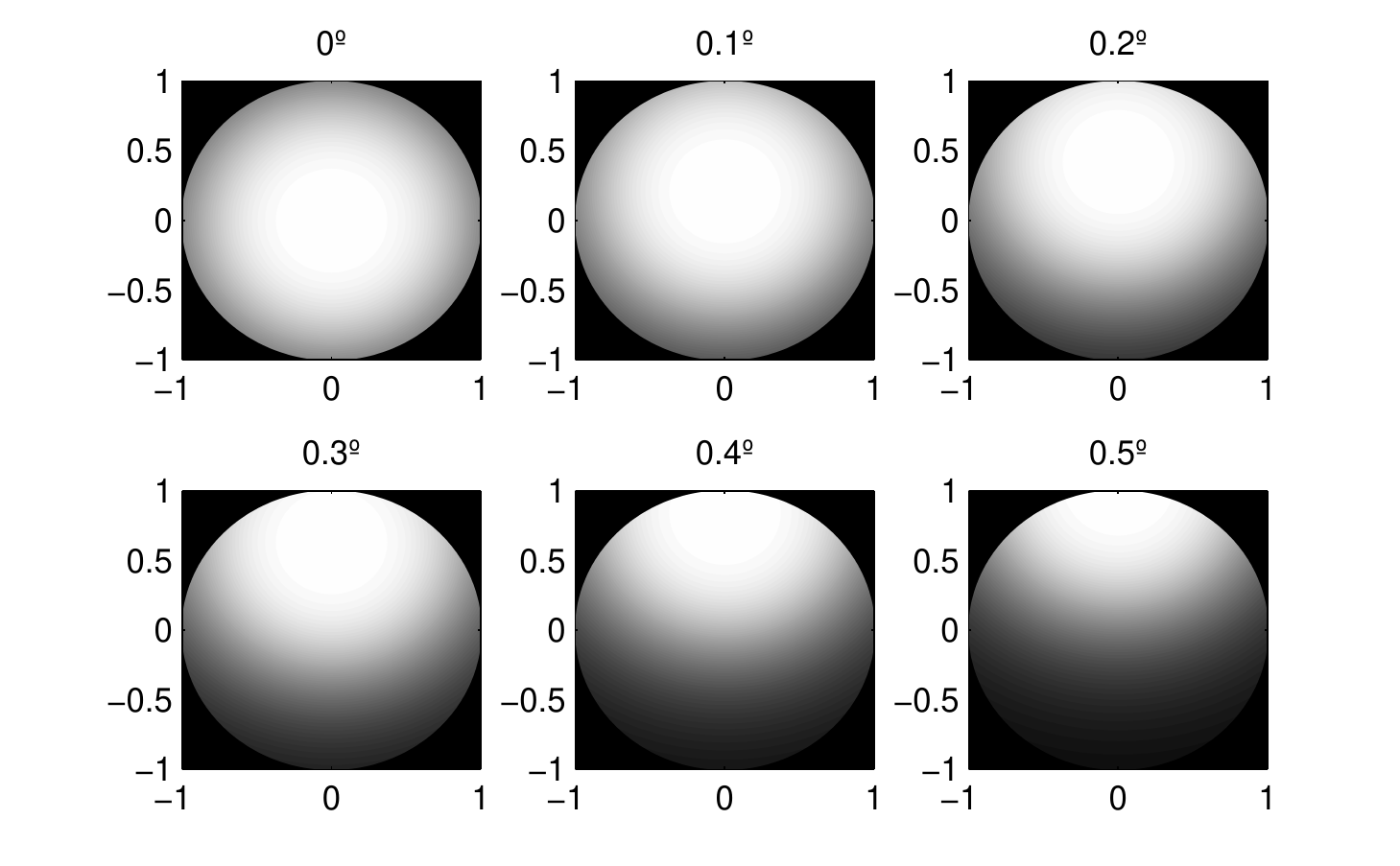}
	\end{center}
	\caption{Apodization of pupil as seen from the etalon for a telecentric beam with $f/60$ and at different angles of incidence of the chief ray in the vertical direction: $\Theta=0\degree$, $0\fdeg1$, $0\fdeg2$, $0\fdeg3$, $0\fdeg4$ and $0\fdeg5$ from the upper left to the lower right. Coordinates have been normalized to the pupil radius.}
	\label{apodization}
\end{figure}

Real instruments cannot strictly follow the requirements for a perfect telecentric system. In an imperfect telecentric instrument, the entrance pupil is not exactly located at the focal plane of the instrument, and the exit pupil is at an intermediate position  between the lens and infinity. The situation is exemplified in Fig. (\ref{telecentric2}) where, without loss of generality, the pupil is assumed at the same location as the lens. The main consequence is that the chief ray cannot be normal to the etalon but is at an incidence angle $\Theta$, which varies across the image. Real instruments always have tolerances for such an incidence angle that cannot be exactly zero as in the ideal case. With such an oblique chief ray, the pupil apodization gets asymmetric. Figure \ref{apodization} displays the pupil illumination as seen from the etalon as a function of the chief ray angle of incidence. While the radial decrease in brightness is symmetric at $\Theta = 0$, it becomes more asymmetric as soon as $\Theta$ increases. The result certainly has an influence in the PSF that varies across the field of view.

\subsection{PSF shape over the FOV}
\begin{figure}
	\begin{center}
		\includegraphics[width=\columnwidth]{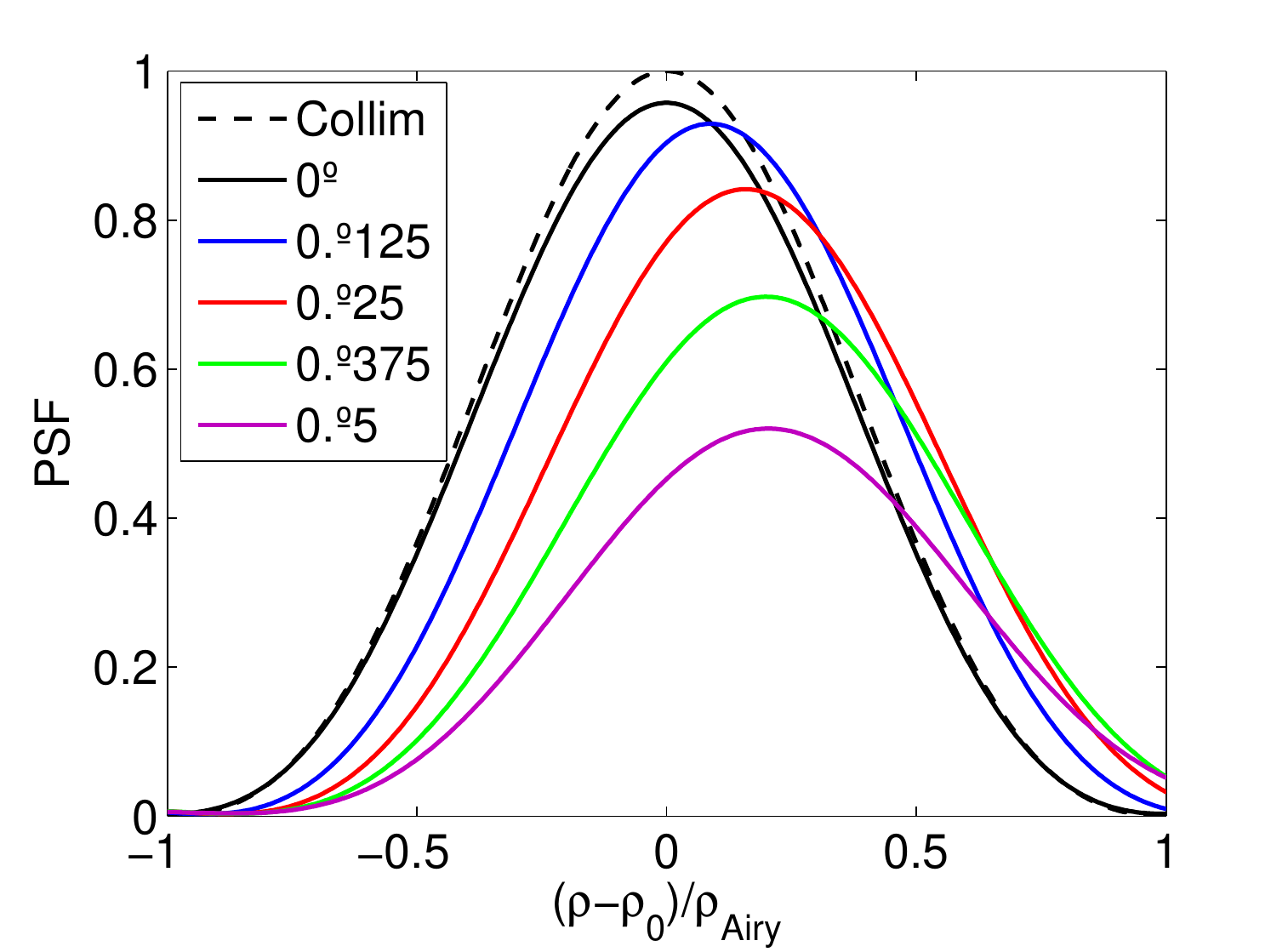}
	\end{center}
	\caption{PSF profiles of the a telecentric etalon with $f/80$ at $\lambda_t$ and at different angles of incidence of the chief ray: $\Theta=0\degree$ (black), $0\fdeg125$ (blue), $0\fdeg25$ (red), $0\fdeg375$ (green) and $0\fdeg5$ (magenta). Each profile is centered at $\rho_0=f\sin\Theta$.}
	\label{PSF_fov}
\end{figure}

Figure \ref{PSF_fov} shows the monochromatic PSF at the peak wavelength at normal incidence, $\lambda_t$, corresponding to a beam with f/80 for different angles of incidence of the chief ray against the radial coordinate of the image plane, $\rho$,  centered at $\rho_0=f\sin\Theta$ (corresponding to the maximum of the PSF of a collimated beam in a circular aperture with incident angle $\Theta$) and normalized by the width of the Airy pattern, $\rho_{\rm Airy}$.  We can observe: (1) a spatial shift of the maximum with respect to the collimated case, (2) a broadening of the PSF, and (3) a decrease of the peak transmitted intensity across the FOV. It is also important to remark that perfect telecentrism is recovered at $\Theta=0$, since $\Theta$ defines in a certain sense the degree of telecentrism. The fact that the PSF is not centered readily implies stray light from the surroundings. Note that $\sim 0.2 \rho_{\rm Airy}$ (the approximate peak of the PSF for $\Theta = 0\fdeg5$) means a third of a pixel in a critically sampled instrument. The broadening of the PSF drives the results in the same direction.

\begin{figure}
	\begin{center}
		\includegraphics[width=\columnwidth]{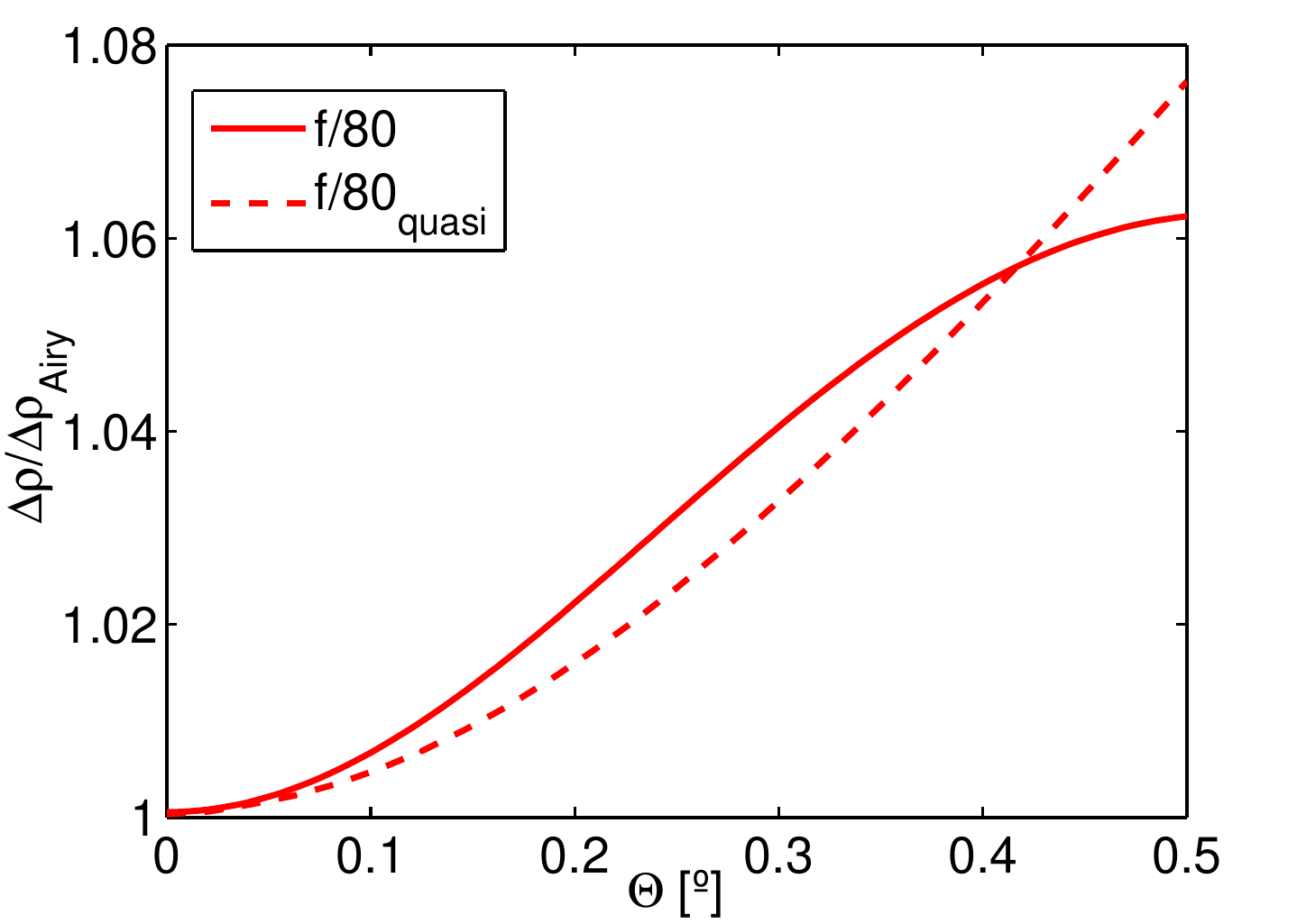}
	\end{center}
	\caption{FWHM of the PSF at $\lambda_t$ normalized by the FWHM of the Airy disk across $\Theta$ for a $f/80$ beam (red solid line). The FWHM of the quasi-monochromatic PSF (red dashed line) and the FWHM of the PSF for a collimated beam have also been plotted (black solid line).}
	\label{PSF_radial}
\end{figure}

\begin{figure}
	\begin{center}
		\includegraphics[width=\columnwidth]{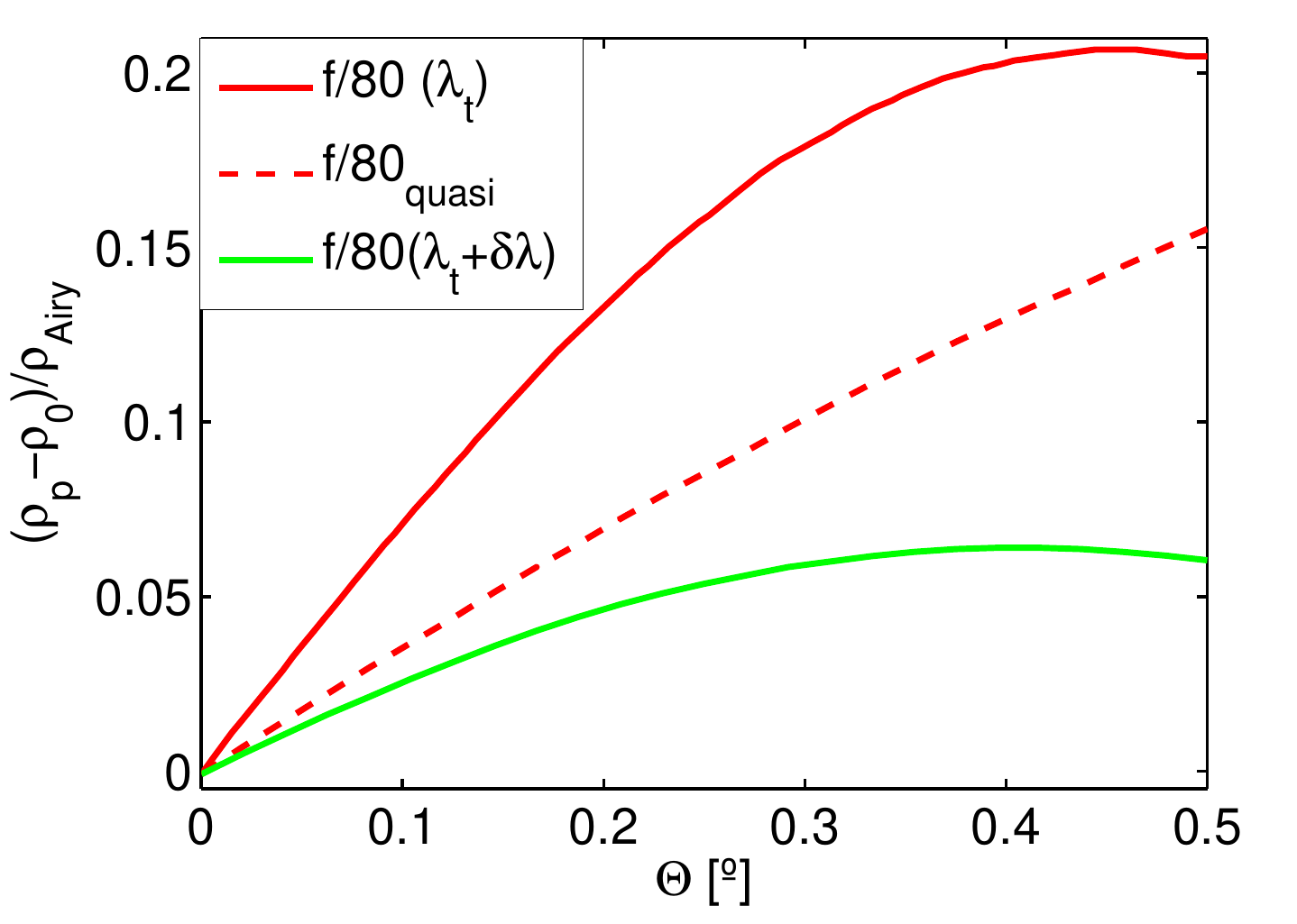}
	\end{center}
	\caption{Spatial shift of the peak of the PSF at $\lambda_t$ with respect to $\rho_0$ across $\Theta$ for a telecentric configuration with $f/80$ (red solid line). The shift for the quasi-monochromatic PSF (red dashed line), as well as at $\lambda+\delta\lambda$ (green solid line) are also represented.}
	\label{shift_FOV}
\end{figure}

The change of the (normalized) FWHM against $\Theta$ is shown in Figure \ref{PSF_radial} for the monochromatic ($\lambda_t$) and quasi-monochromatic PSFs of an imperfect telecentric configuration illuminated with an f/80 beam. It is to be noticed that the PSF width grows monotonically with the chief ray incidence angle. The variation of width at $0\fdeg5$ is about 7\%  and $8\%$ for the monochromatic and quasi-monochromatic curves respectively.
Figure \ref{shift_FOV} shows the spatial shift of the PSF peak, $\rho_p$, with respect to $\rho_0$ against  $\Theta$ for $\lambda_t$, $\lambda_t+\delta\lambda$ ($\delta\lambda=5$ pm) and for the quasi-monochromatic PSF. The etalon is illuminated with a $f/80$ beam in all cases. The spatial displacement is about 18$\%$ and $15\%$ at 0.5$º$ for the monochromatic PSF at $\lambda_t$ and for the quasi-monochromatic PSF respectively. Interestingly, the dependence at $\lambda_t+\delta\lambda$ is different from that at $\lambda_t$, which indicates that the shift is wavelength dependent and that the PSFs overlap not only spatially but also spectrally over the image plane. 

\subsection{Behavior of the spectral profile over the FOV}
\begin{figure} [t]
	\begin{center}
		\includegraphics[width=\columnwidth]{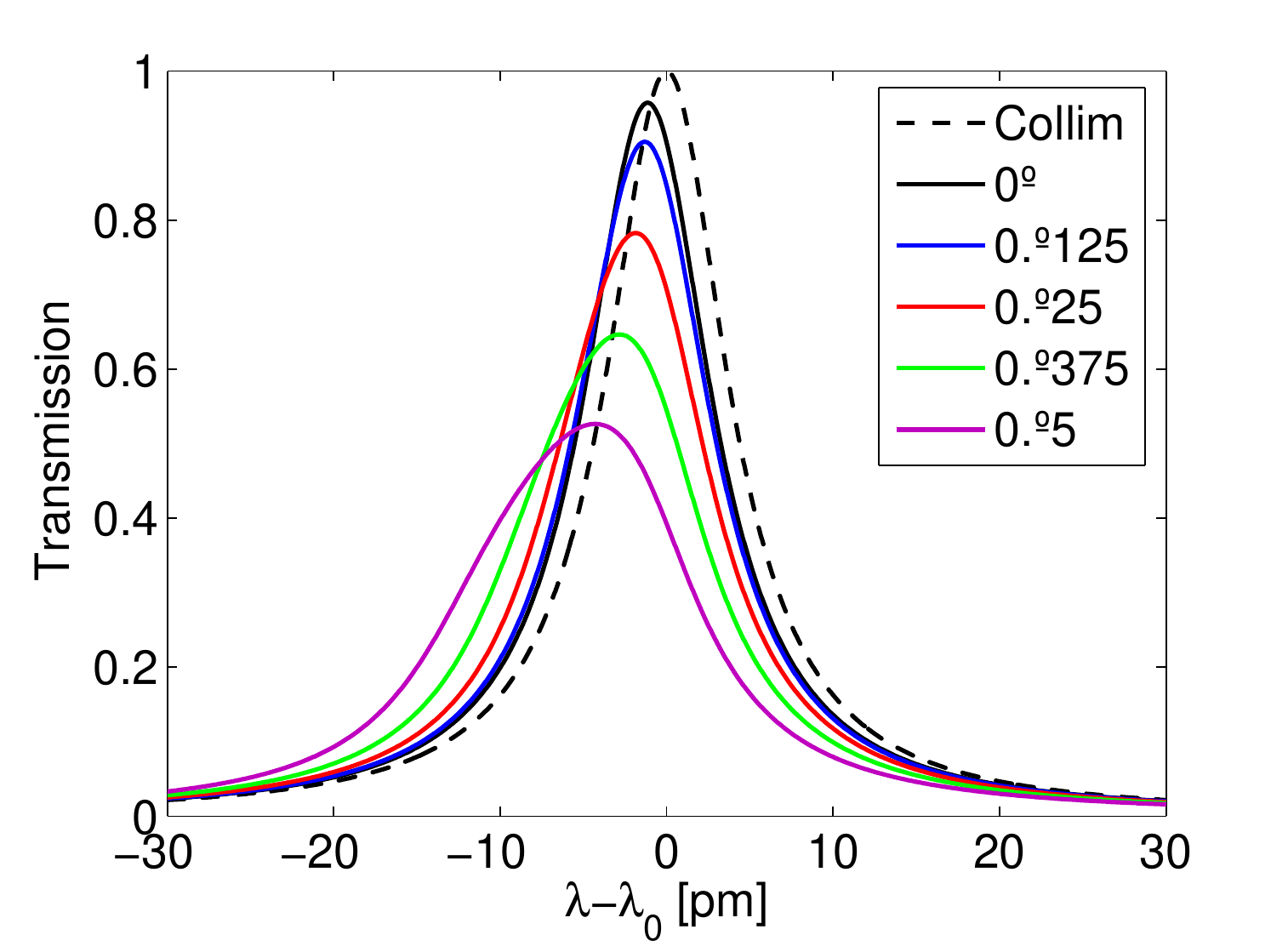}
		\caption{Spectral transmission function at $f/60$ for chief ray angles of incidence $\Theta=0\degree$ (black), $0\fdeg125$ (blue), $0\fdeg25$ (red), $0.375$ (green), and $0\fdeg5$ (magenta).}
		\label{transmission_fov}
	\end{center}
\end{figure}

The loss of symmetry in the cone of rays is also mapped into the transmission profiles of the etalon. These profiles will be shifted and deformed, as happens with the PSFs. Figure \ref{transmission_fov} shows the transmission profile as a function of the wavelength distance to $\lambda_0$ for $\Theta=0\degree$ , $0\fdeg125$, $0\fdeg25$, $0\fdeg375$ and $0\fdeg5$. A beam with f/60 has been employed to clearly visualize the asymmetrization and loss of illumination with $\Theta$. We can appreciate the blue shift across the FOV, as well as a decrease of the symmetry, a broadening of the profiles and a decrease of the peak transmitted intensity as $\Theta$ grows. Also note that at $\Theta=0$ we recover the transmission profile for f/60 presented in Figure \ref{fig:transcoltel}. 

\begin{figure}[t]
	\begin{center}
		\includegraphics[width=\columnwidth]{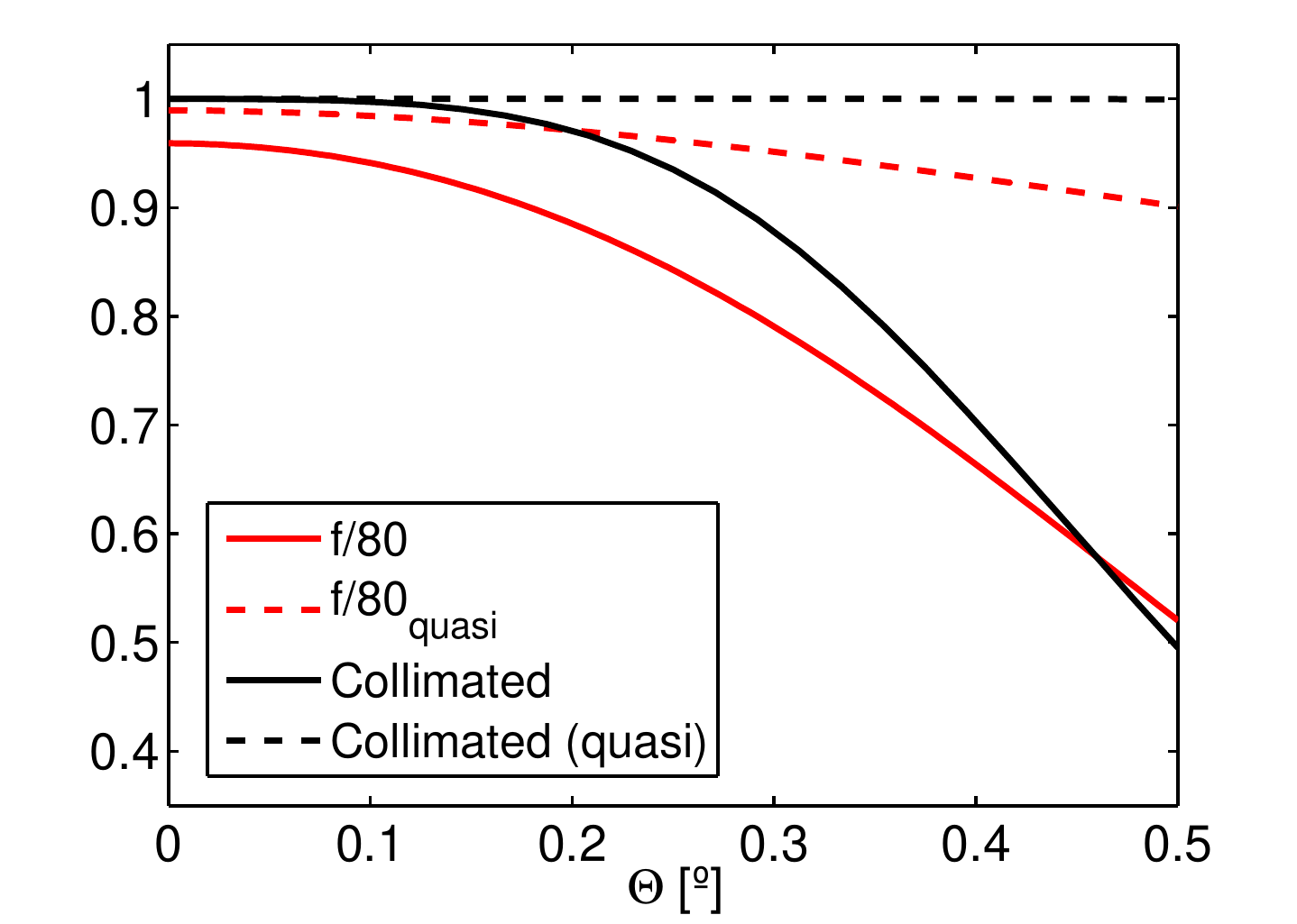}
		\caption{Transmitted intensity at $\lambda_t$ for a telecentric configuration with $f/80$ across $\Theta$ (red solid line) and for a collimated configuration (black solid line). The total flux of energy transmitted normalized by the flux transmitted in the collimated configuration (black dashed line) is also represented (red dashed line).}
		\label{I_FOV}
	\end{center}
\end{figure}

Figure \ref{I_FOV} shows the transmitted intensity with $\Theta$ evaluated at the wavelength of the peak transmission for normal illumination, $\lambda_t$. A beam with f/80 has been employed. The decay of transmission  at $\lambda_0$ with the incident angle of the collimated case is also represented. The peak intensity goes from $0.96$ and $1$ at $0\degree$ to $0.52$ and $0.49$ at $0\fdeg5$ for the telecentric and collimated beams respectively. It should be noticed that in the collimated case the intensity decays faster with the incidence angle. 
 We also show the total energy contained in the transmission profiles for both the telecentric and collimated beams with the chief ray incidence angle. We have normalized both to the total energy contained in the transmission profile of the collimated configuration (which remains constant over $\Theta$). The total energy of the profile is calculated by integrating the spectral transmission factor, $\tilde{g}$. We can observe that the flux of the telecentric configuration is reduced about $9\%$ from the center of the image to its edges.

\begin{figure} [t]
	\begin{center}
		\includegraphics[width=\columnwidth]{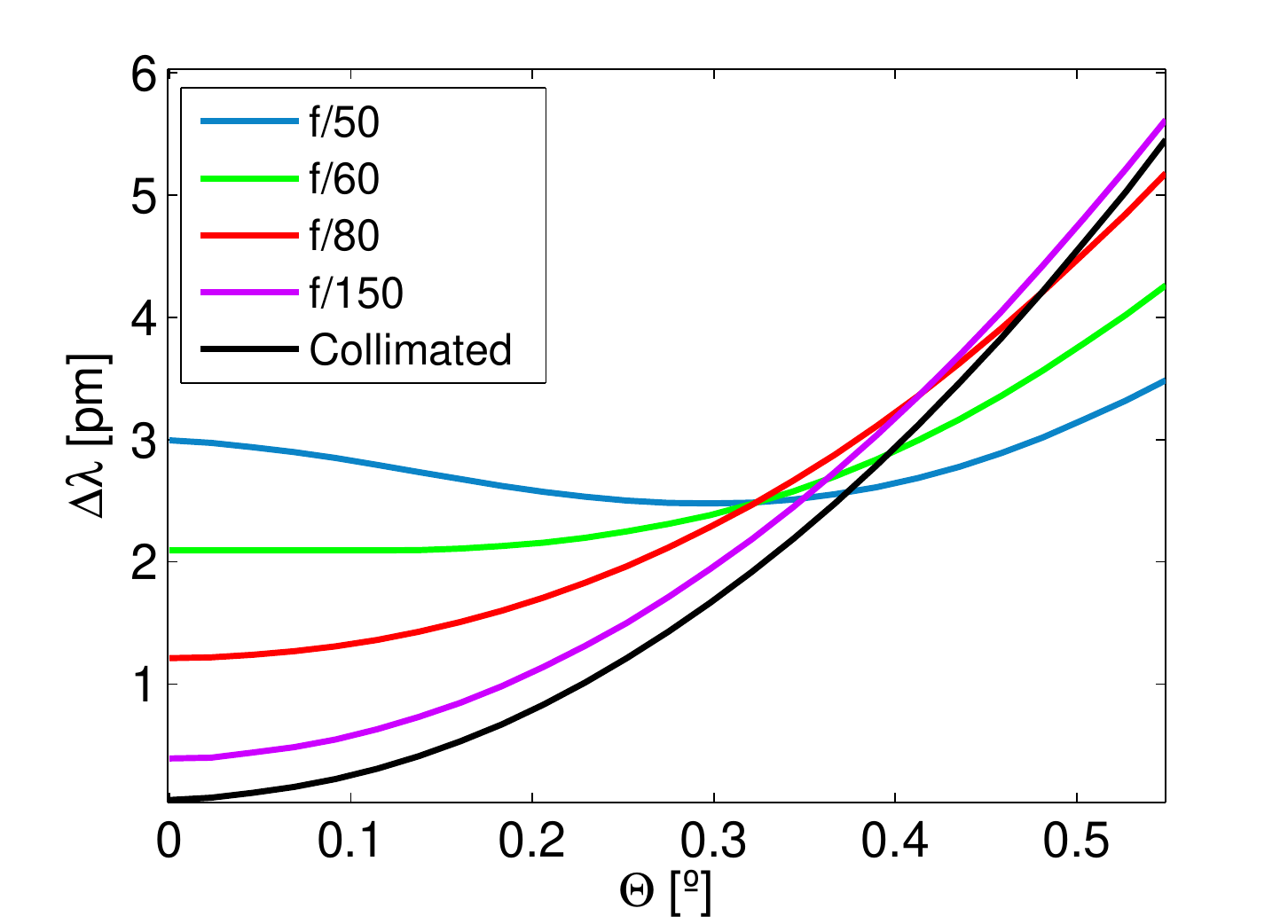}
		\caption{Spectral shift of the peak wavelength across $\Theta$ for f-numbers $f/50$ (blue), $f/60$ (green), $f/80$ (red), and $f/150$ (magenta). As reference, the spectral shift of the collimated configuration is also plotted (black line). }
		\label{spectral_fov}
	\end{center}
\end{figure}

Although the telecentric configuration was devised to avoid the wavelength shift, $\Delta\lambda$, across the FOV, characteristic to the collimated configuration, a wavelength shift will appear in real instruments, as seen in Figure \ref{transmission_fov}. Figure \ref{spectral_fov} shows the spectral displacement of the wavelength peak with $\Theta$ for the nominal wavelength, $\lambda_0$, for different f-numbers. 
 It can be noticed that the shift goes towards the blue for all angles and has a weaker dependence on $\Theta$ as the f-number decreases from infinity (collimated case) to $\simeq 60$. For $f\#<60$, the shift is reduced as $\Theta$ increases until reaching a minimum at a certain value (larger for smaller f-numbers) and then grows monotonically towards the blue. The weaker dependence with smaller f-numbers contrasts with other effects, such as the broadening and the asymmetrization of the PSF and of the spectral profile, where the effect is more prominent for smaller f-numbers. This indicates that a compromise must be reached in general between the spectral shift and the degradation of the PSF and of the spectral transmission with the f-number in our instruments.

To qualitatively understand why the wavelength shift decreases or increases over the FOV depending on both the f-number and the chief ray angle, let us take a look to Figure \ref{telecentric2}. If we set $\Theta=0$ (normal illumination of the pupil), the cone of rays becomes symmetric and the maximum incidence angle is the same at both sides of the optical axis. The effect is a wavelength displacement of the peak wavelength towards the blue of the nominal wavelength, $\lambda_0$. As $\Theta$ increases, the maximum incidence angle decreases at one side of the optical axis and increases at the other side. This causes a sort of trade-off to increase or to decrease the shift with respect to normal illumination when averaging the electric field transmitted by the etalon over the pupil. This is of course $f\#$ dependent as  the cone of rays reaching the etalon narrows when the f-number increases and vice-versa.

\section{Summary and conclusions}
We have discussed the properties of Fabry-P\'erot etalons in the two optical configurations commonly employed in solar instruments, namely, collimated and telecentric. We have focused on both their use as tunable spectral filters and as imaging elements.

First, we have overviewed the general properties of Fabry-P\'erot etalons, their tunability and their sensitivity to variations in the optical thickness. We have remarked that changes in the optical thickness specially affect the peak wavelength but not so much to the shape of the transmission profile.

 We have studied the degradation of the spectral profiles originated by both etalon defects and illumination with a beam of a certain aperture. We have followed the general treatment given by \cite{sloggett} and we have extended their results by presenting explicit formulas for the finesse defects of typical inhomogeneities (spherical, Gaussian, parallelism and sinusoidal). The found expressions are valid for irregularities having a small effect in the transmission profile.
 We have also obtained formulas for the finesse defects in the opposite limit, i.e., when irregularities dominate. We have shown that these finesses agree with the limiting formulas of \cite{chabbal}, commonly employed in the literature but only valid for defects that produce a severe degradation of the profile. They differ from the small defect case for the Gaussian and sinusoidal defects, whereas they coincide for the spherical and parallelism error, as expected. 
On the other hand, we have generalized the aperture finesse presented by \cite{ref:atherton}, to the crystalline case and we have deduced an analytical expression for the blue shift of a telecentric etalon. The derived expressions show a good agreement with results obtained from numerical simulations of the spectral profile.

Regarding their imaging performance, we have shown that, in a collimated mount, the PSF is proportional to that of an ideal diffraction-limited instrument. The proportionality factor is given by the spectral profile of the etalon. This implies that convolution with the object cannot be applied since the PSF is not space invariant. A monochromatic decrease of the S/N is then expected from the center to the borders of the image. However, the decay of monochromatic transmission can be dramatic if variations of the optical thickness do not preserve the peak shift low compared to the width of the profile. In a perfect telecentric etalon the PSF remains the same from point to point but strongly depends on the wavelength over the transmission profile. This gives rise to artificial velocities and magnetic fields that can only be calibrated in a first approximation.

 We have argued that fluctuations of the optical path due to defects are averaged in collimated setups and only affect at first order to the transmission as we go off axis. Stray-light is also expected if micro-roughness errors are present. The PSF shape remains equal and symmetric all over the FOV, though.  In a telecentric setup, imperfections in the optical path produce a change on the PSF pixel-to-pixel and further contribute to artificial velocity and magnetic field signals. In the case that two or more etalons are combined to increase the effective free spectral range and/or to improve the  resolution, the errors can be amplified in both mounts, and may also produce large local transmission variations in the telecentric configuration because of the different shifts of the spectral profiles due to different local thicknesses. 
 
 We have added in our discussion the effect of the quasi-monochromatic nature of the measurements due to the finite passband of the etalons. The response of the instrument turns out to depend, in general, on the object itself in both the collimated and the telecentric configurations. Therefore, the quasi-monochromatic PSF cannot be employed as a regular one, except for observations of spectrally flat features (i.e., in the continuum). Purely monochromatic effects, such as the decay of intensity in collimated etalons and the artificial signals originated in telecentric ones, are expected to balance out in some way, although not entirely. Quantitative effects can only be evaluated numerically.
  
We have finally addressed the consequences of variations on the chief ray over the FOV in telecentric setups. We have shown that they can produce a severe asymmetrization, a broadening and a shift of both the peak transmission and the PSF. These effects are nonlinear with the angle and with the f-number, and, thus, very sensitive with the optical tolerances of the instrument. A decrease of the transmitted flux of photons with larger incidence angles of the chief ray has also been demonstrated, apart from a reduction the monochromatic intensity. All these issues, except for the shift of the peak, are not present in the collimated configuration and, when combined, lead to artificial signals in the spectrum of the measured Stokes vector and to a degradation of the image. Also, the widening and shift of the spectral profile in imperfect (real) telecentric mounts contradicts the general conception of employing this configuration to keep the passband constant over the field of view. 

The consequences of imperfect telecentrism can also be applied to imperfect collimated mounts, where the etalon is illuminated by a finite f-number beam with different incidence angles on the etalon. The only difference is that defects still average over the footprint of the beam on the etalon. 

\begin{acknowledgements}
This study was initiated upon some starting notes by our friend Jos\'e Antonio Bonet, a colleague for most of the development phases of the IMaX and SO/PHI instruments. We owe very much to these notes (including a couple of figures) and would like to publicly (and warmly) thank his contribution. This work has been supported by Spanish Ministry of Economy and Competitiveness through projects ESP2014-56169-C6-1-R and ESP-2016-77548-C5-1-R.  DOS also acknowledges financial support through the Ram\'on y Cajal fellowship.
\end{acknowledgements}

\appendix
\section{A: Derivation of the small-defect finesse expressions}
\label{appendix}
To model real etalons with typical defects, we call $\epsilon$  the error in $\delta$ induced by an optical path defect $\Delta s$. Thus,
\begin{equation}
\label{eq:tildedelta}
\epsilon = \frac{4\pi\Delta s}{\lambda}=\frac{\Delta s}{s} \, \delta,
\end{equation}
where $\delta$ is given by Equation (\ref{eq:phasedif}), $\lambda$ is the wavelength of the incident beam and $s\equiv n\prima h\cos\theta\prima$ is the optical path in one pass of the beam through the etalon. Let $D(\epsilon)$ be the probability density function for $\epsilon$, so that $D(\epsilon_{0}) \, \de$ is the surface fraction of the etalon aperture, $\dS$, for which the error in $\delta$ is in the range $(\epsilon_{0},\epsilon_{0}+{\rm d}\epsilon)$. That is:

\begin{equation}
\dS=D(\epsilon){\rm d}\epsilon,
\end{equation}
where
\begin{equation}
\de=\frac{4\pi}{\lambda}{\rm d}s.
\label{de}
\end{equation}
By definition, the error distribution function is just given then by

\begin{equation}
D(\epsilon)=\kappa\frac{{\rm d}S}{\de},
\label{density}
\end{equation}
where $\kappa$ is a normalization factor introduced for $D(\epsilon)$ to represent a probability density function in a strict sense, that is, to fulfill the property

\begin{equation}
\int_{-\infty}^{\infty}\!\! D(\epsilon)\de=1.
\end{equation}

Let us call $\mu_d$ and $\sigma_d^2$ the mean and variance of the distribution respectively. Assuming defects are small ($\alpha=2\sqrt{3}$),  Eq. (\ref{eq:defectfinesse}) can be expressed then as \citep{sloggett}

\begin{equation}
\label{eq:smalldefect}
{\cal F}_d=\frac{\pi}{\sigma_d\sqrt{3}}.
\end{equation}

By relating the variance of the defect distribution with measurable parameters of the defect, such as departure from an ideal flat surface, we can obtain useful expressions for the defect finesse.
\subsection{Spherical defect}
We will focus first on the spherical-shape defect shown in Figure \ref{fig:platedefects} (a). If we consider an etalon with circular or parabolic symmetry and define $r$ as the radial coordinate (Fig. \ref{fig:defect_coordinates}a), the optical path across the etalon surface is given by
 
 \begin{equation}
 s=ar^2+s_0,
 \end{equation}
with a peak-to-peak excursion $\Delta s_{\rm s}=aR^2$, where $a$ is a proportionality factor, $R$ is the radius of the etalon, and $s_0$ is the optical path at $r=0$. The differential of the optical path can be expressed just as
 \begin{equation}
 {\rm d}s=2ar\dr.
 \end{equation}
 Therefore, Eq. (\ref{de}), can be cast as
  \begin{equation}
 \de=\frac{8\pi}{\lambda}ar\dr.
 \end{equation}
On the other hand, taking advantage of the symmetry of the problem,

\begin{equation}
\dS=2\pi r\dr.
\end{equation}
Substituting this expression in Eq. (\ref{density}) we have that $D_{\rm s}$ is a rectangular distribution that, after normalization ($\kappa=4a\lambda^{-1}\epsilon_{\rm max}^{-1}$), can be written as
\begin{equation}
\label{eq:disspherical}
D_{\rm s}(\epsilon)= \left\{ \begin{array}{ll}
\epsilon_{\rm max}^{-1} & {\rm if} \,\,\,\, 0<\epsilon\leq \epsilon_{\rm max} \\ \\
0 & {\rm otherwise} 
\end{array} \right.,
\end{equation}
 where $\epsilon_{\rm max} = (4\pi /\lambda) \Delta s_{\rm s}$. This distribution is also useful for uniformly distributed random defects and for aperture defects \citep{sloggett}. Its mean is just $\mu_{d_{\rm s}}=\epsilon_{\rm max}/2$ and its variance is given then by 
\begin{equation}
\sigma_{d_{\rm s}}^2=\frac{1}{12}\epsilon_{\rm max}^{2}.
\label{eq:sigma_ds}
\end{equation}

Consequently, substituting $\sigma_{d_{\rm s}}$ in Eq. (\ref{eq:smalldefect}), we get that
\begin{equation}
{\cal F}_{d_{\rm s}}=\frac{\lambda}{2\Delta s_{\rm s}}.
\end{equation}

Note that $\epsilon_{\rm max}$ also coincides with the FWHM of the defect distribution in this particular case. Therefore, the relation between $\sigma_{d_{\rm s}}$ and the FWHM is  given by $w_{d_{\rm s}}=2\sqrt{3}\sigma_{d_{\rm s}}$ and the limiting finesse is expected to coincide with that here deduced for small defects.
\subsection{Gaussian random defect}
If we now consider Fig.\ \ref{fig:platedefects} (b), we have a micro-rough surface with deviations from $s$ that follow a normalized Gaussian distribution with variance $\Delta s_{\rm g}^{2}$. In this case, the standard deviation of the distribution is obviously $\sigma_{d_{\rm g}} = 4\pi\, \Delta s_{\rm g}/\lambda$. Substituting this value in Eq. (\ref{eq:smalldefect}),

\begin{equation}
{\cal F}_{d_{\rm g}}=\frac{\lambda}{4\sqrt{3} \Delta s_{\rm g}}.
\label{eq:ap_gauss}
\end{equation}

On the other hand, the FWHM of this distribution is related to the standard deviation as $w_{d_{\rm g}}=2\sqrt{2\ln2}\sigma_{d_{\rm g}}$. Therefore, the value of $\alpha$ will tend to $2\sqrt{2\ln2}\sigma_{d_{\rm g}}$ when defects dominate  and the limiting finesse will differ from the finesse here deduced for small defects.

\subsection{Parallelism defect}
\begin{figure}[ht]
	\begin{center}
		\includegraphics[width=0.7\textwidth]{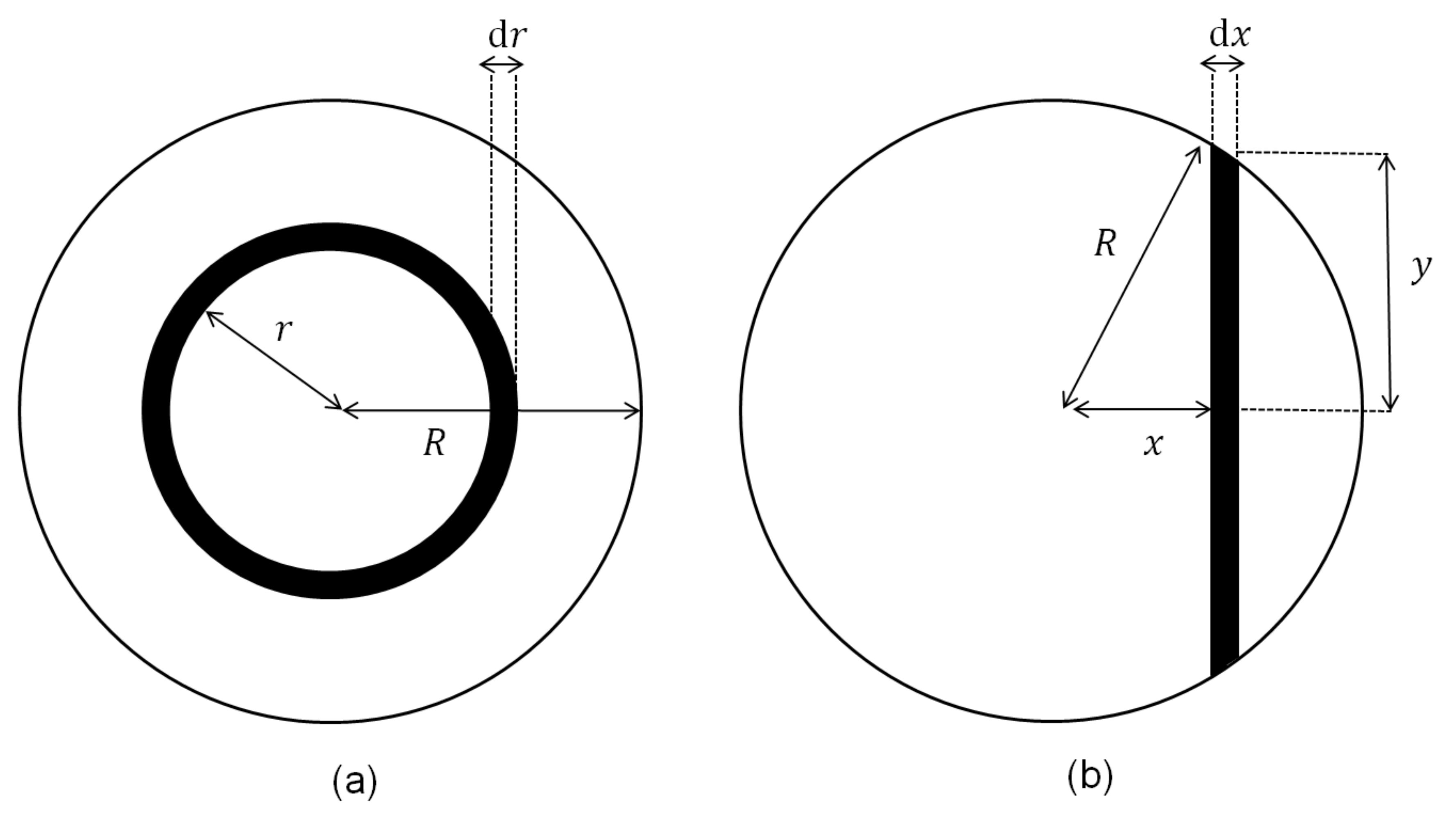}
	\end{center}
	\caption{Coordinates employed for the calculation of the density distribution function of different defects: (a) coordinates used in in the spherical defect; (b) coordinates used in both the parallelism and sinusoidal defect.}
	\label{fig:defect_coordinates}
\end{figure}

For the parallelism defect shown in Figure \ref{fig:platedefects} (c), if we consider a circular etalon with radius $R$ and define the $X$ direction to be the direction of departure from parallelism (Figure \ref{fig:defect_coordinates}b), the optical path depends on the $x$ coordinate as
\begin{equation}
s=ax+s_0,
\end{equation}
with a peak-to-peak deviation from paralellism $ \Delta s_{\rm p}=2aR$, where $a$ is a proportionality factor and $s_0$ is the optical path at $x=0$. The differential of the optical path is simply
\begin{equation}
{\rm d}s=a\dx.
\end{equation}
Therefore, using Eq. (\ref{de})
\begin{equation}
\de=\frac{4\pi a}{\lambda}\dx.
\end{equation}
On the other hand,
\begin{equation}
\dS=2 y\dx=2(R^2-x^2)^{1/2}\dx,
\end{equation}
and, substituting in Eq. (\ref{eq:tildedelta}) we have that
\begin{equation}
x=\frac{\epsilon\lambda}{4\pi a}.
\end{equation}
Replacing $x$ in Eq. (\ref{density}) and restricting to $|\epsilon| \leq \epsilon_{\rm max}/2$, where $\epsilon_{\rm max} = (4\pi /\lambda) \Delta s_{\rm p}$, we can express
\begin{equation}
D_{\rm p}(\epsilon)=\kappa\frac{\lambda}{2\pi}\left[R^2-\left(\frac{\lambda}{4\pi a}\right)^2\epsilon^2\right]^{1/2}.
\end{equation}
The normalization constant is given in this case by $\kappa=(\pi a R^2)^{-1}$. We can cast this equation more elegantly as
\begin{equation}
\label{eq:parallelism}
D_{\rm p}(\epsilon)= \left\{ \begin{array}{ll}
4/(\pi\epsilon_{\rm max}) \left[ 1-4\epsilon^2/\epsilon_{\rm max}^2 \right]^{1/2} & {\rm if} \,\,\,\, |\epsilon| \leq \epsilon_{\rm max}/2 \\ \\
0 & {\rm otherwise} 
\end{array} \right.,
\end{equation}
Note that the mean is zero as it is symmetrical about $\epsilon=0$. The variance is given by

\begin{equation}
\sigma_{d_{\rm p}}^2=\frac{\epsilon_{\rm max}^2}{16}.
\end{equation}
Then, $\sigma_{d_{\rm p}} = \pi \Delta s_{p} / \lambda$ and the defect finesse (Eq. \ref{eq:smalldefect}) can be written as

\begin{equation}
{\cal F}_{d_{\rm p}}=\frac{\lambda}{\sqrt{3}\Delta s_{\rm p}}.
\end{equation}

The FWHM of this distribution is given by $w_{d_{\rm p}}=2^{-1}\sqrt{3}\epsilon_{\rm max}=2\sqrt{3}\sigma_{d_{\rm p}}$. As for the spherical defect, the limiting value of $\alpha$ is expected then to tend to $2\sqrt{3}$ for large defects. 

\subsection{Sinusoidal defect}
Consider finally an etalon with an optical path roughness given by a sinusoid of peak-to-peak amplitude $\Delta s_{\rm a}$, as shown in Figure \ref{fig:platedefects} (d). If we consider a circular etalon with radius $R$ and define the $X$ direction to be the direction of the sinusoid, the optical path have a dependence with the $x$ coordinate
\begin{equation}
s=\frac{\Delta s_{\rm a}}{2}\sin \omega x+s_0,
\end{equation}
where $\Delta s_{\rm a}$ is peak-to-peak deviation amplitude of the sinusoid, $\omega$ is its (spatial) frequency of oscillation, and $s_0$ is the optical path at $x=0$. The differential of the optical path is given by
\begin{equation}
{\rm d}s=\frac{\Delta s_{\rm a}}{2}\omega\cos\omega x\,\,\dx.
\end{equation}
Therefore, using Eq. (\ref{de})
\begin{equation}
\de=\frac{2\pi \omega}{\lambda}\Delta s_{\rm a}\sqrt{1-4(s-s_0)^2/\Delta s_{\rm a}^2}\ \dx,
\end{equation}
and, according to Eq. (\ref{eq:tildedelta}),
\begin{equation}
\de=\frac{2\pi \omega}{\lambda}\Delta s_{\rm a}\sqrt{1-(\epsilon\lambda)^2/(2\pi\Delta s_{\rm a})^2}\dx.
\end{equation}
On the other hand, for a circular etalon
\begin{equation}
\dS=2 y\dx=2(R^2-x^2)^{1/2}\dx,
\end{equation}
where $x$ is related to $\epsilon$ by
\begin{equation}
x=\frac{1}{\omega}\left[\arcsin{\left(\frac{2(s-s_0)}{\Delta s_{\rm a}}\right)}+ 2\pi n\right].
\end{equation}
The term $2\pi n$, accounts for the multiplicity of the solutions,  where $n=0,\pm1,\pm2,...,\pm N$, and
\begin{equation}
N=\frac{1}{2\pi}\left[\omega R-\arcsin\left(\frac{2(s-s_0)}{\Delta s_{\rm a}}\right)\right].
\end{equation}
Then, in the range  $|\epsilon| \leq \epsilon_{\rm max}/2$, with $\epsilon_{\rm max} = (4\pi /\lambda) \Delta s_{\rm a}$,

\begin{equation}
D(\epsilon)\simeq\kappa\frac{\lambda\sqrt{R^2-\omega^{-2}\left(\arcsin\left[2(s-s_0)/\Delta s_{\rm a}\right]+2\pi n\right)^2}}{\pi \omega\Delta s_{\rm a}\left[1-(\epsilon\lambda)^2/(2\pi\Delta s_{\rm a})^2\right]^{1/2}}.
\end{equation}
If we approximate $R^2>\!\!>x^2$, which is valid for fast spatial modulations of the sinusoidal defect and for $n<\!\!<N$ 
\begin{equation}
D(\epsilon)\simeq\kappa\frac{R\lambda}{\pi \omega\Delta s_{\rm a}\sqrt{1-(\epsilon\lambda)^2/(2\pi\Delta s_{\rm a})^2}}=\kappa\frac{2R}{\omega\sqrt{\epsilon_{\rm max}^2/4-\epsilon^2}},
\end{equation}
where the normalization constant can be shown to be given by $\kappa=\omega(2R\pi)^{-1}$. The probability density function can then be cast as
\begin{equation}
\label{eq:sinusoid}
D_{\rm a}(\epsilon)= \left\{ \begin{array}{ll}
\pi^{-1} [\epsilon_{\rm max}^2/4 - \epsilon^{2}]^{-1/2} & {\rm if} \,\,\,\, |\epsilon| \leq \epsilon_{\rm max}/2 \\ \\
0 & {\rm otherwise} 
\end{array} \right.,
\end{equation}
Due to the symmetry of the distribution, $\mu_{\rm a}=0$ and

\begin{equation}
\sigma_{d_{\rm a}}^2=\frac{\epsilon_{\rm max}^2}{8}.
\end{equation}
Then, the finesse defect can be expressed as (Eq.\ref{eq:smalldefect})

\begin{equation}
{\cal F}_{d_{\rm a}}=\frac{\lambda}{\Delta s_{\rm a}\sqrt{6}}.
\end{equation}

The FWHM of this distribution, $w_{d_{\rm a}}$, is just $\epsilon_{\rm max}$. Therefore, $w_{d_{\rm a}}=2\sqrt{2}\sigma\simeq2.83\sigma_{d_{\rm a}}$. The value of $\alpha$ will tend then to $2\sqrt{2}$ when defects dominate and the limiting finesse will differ from the one here deduced for small defects.
\subsection{Aperture finesse and spectral shift in telecentric configuration}
Following the arguments of \cite{sloggett}, we can also deduce an expression for the aperture finesse. Let us consider that the etalon with refraction index $n\prima$ is at the focal plane of a lens of radius $R$. Each point of the etalon will receive rays coming from all parts of the lens. In this case, the phase error corresponding to each ray with incidence angle $\theta$ from a medium with refraction index $n$, compared to normal incidence and for $\theta<\!\!<1$, is given by

\begin{equation}
\epsilon=\frac{4\pi\nprima h}{\lambda}(1-\cos\theta')\simeq \frac{n^2 m \pi \theta^2}{\nprimac},
\label{epsilonaperture}
\end{equation}
where $m=2n\prima h\lambda^{-1}$ is the interferential order for $\theta\simeq0$. First, we shall calculate the density distribution of the incidence angle $\theta$ in the etalon, $D(\theta)$. Similarly to Eq. (\ref{density})

\begin{equation}
D(\theta)=\kappa \frac{\dS}{{\rm d}\theta}.
\end{equation}
Here $\dS$ represents the portion of the lens corresponding to the angles of the rays coming to the etalon with angles between $(\theta_0,\theta_0+{\rm d}\theta)$. Let $r$ be the radial coordinate of the lens. Then

\begin{equation}
\dS=2\pi r\dr.
\end{equation}
For small angles, $r\simeq\theta f$ and $\dr\simeq\theta f^2{\rm d}\theta$, where $f$ is the focal length of the system. Thus
\begin{equation}
\dS=2\pi f^2\theta{\rm d}\theta,
\end{equation}
and, for $\theta$ in the range $(0,\theta_{\rm m})$, where  $\theta_{\rm m}$ is the maximum incidence angle, the angular distribution is simply
\begin{equation}
D(\theta)= 2\pi\kappa f^2\theta,
\end{equation}
where $\kappa=(\pi f^2\theta_{\rm m}^2)^{-1}$ after normalization. To obtain $D(\epsilon)$ we can use the relation
\begin{equation}
D(\theta){\rm d}\theta=D(\epsilon){\rm d}\epsilon
\end{equation}
and
\begin{equation}
\frac{{\rm d}\theta}{\de}=\frac{\nprimac}{2\pi n^2 m\theta}.
\end{equation}
Then,

\begin{equation}
D_{\rm f}(\epsilon)=\begin{cases}
\nprimac(n^2m\pi\theta_{\rm m}^2)^{-1}       & \text{if}\ \ 0<\epsilon<\epsilon_{\rm max} \\
0  & \quad \text{otherwise}\\
\end{cases},
\label{eq:Df}
\end{equation}
where $\epsilon_{\rm max}=n^2m\pi\theta_{\rm m}^2n^{-2}$. Notice that the density distribution is a rectangular function, as for the spherical defect. Actually, we can rewrite Eq. (\ref{eq:Df}) as 

\begin{equation}
D_{\rm f}(\epsilon)=\begin{cases}
1/(2\sqrt{3}\sigma_{d_{\rm f}})       & 0<\epsilon<2\sqrt{3}\sigma_{d_{\rm f}}\\
0  & \quad \text{otherwise}\\
\end{cases},
\end{equation}
which only differs from Eq. (\ref{eq:disspherical}) by a constant that is not relevant as both distributions are normalized. The mean value of the distribution and its variance turn out to be

\begin{equation}
\mu_{d_{\rm f}}=\frac{\epsilon_{\rm max}}{2}=\frac{1}{2}\frac{n^2m\pi\theta_{\rm m}^2}{\nprimac},
\label{mu_a}
\end{equation}

\begin{equation}
\sigma_{d_{\rm f}}^2=\frac{\epsilon_{\rm max}^2}{12}=\frac{1}{12}\left(\frac{n^2m\pi\theta_{\rm m}^2}{\nprimac}\right)^2,
\end{equation}

Using Eq. (\ref{eq:defectfinesse}) and $\Omega=\pi\theta_{\rm m}^2$, the aperture finesse can be deduced to be (Eq. \ref{eq:smalldefect})

\begin{equation}
{\cal F}_{d_{\rm f}} = \frac{2\pi}{m \Omega}\frac{\nprimac}{n^2}.
\label{F}
\end{equation}

The FWHM of this distribution is therefore also given by $w_{d_{\rm f}}=2\sqrt{3}\sigma_{d_{\rm f}}$. Since $\alpha$ tends for $2\sqrt{3}$ for both the small and large defect regime, the limiting finesse defect will coincide with this expression.

Since the mean value of this distribution is not zero (Eq. \ref{mu_a}), the profile is expected to shift towards the blue. As the density distribution is symmetrical about its mean,  the retardance corresponding to the peak wavelength in telecentric configuration, $\lambda_t$, will be related to the retardance at the peak wavelength for collimated illumination by
 \begin{equation}
 \delta(\lambda_t)=\delta(\lambda_0)+\mu_{d_{\rm f}}(\lambda_0).
 \label{eq:deltalambdat}
 \end{equation}
If we relate the maximum incidence angle with the f-number of the incident beam through $\theta_{\rm m}=(2f\#)^{-1}$, the transmission peak then depends on $\lambda_0$ as
 
 \begin{equation}
\lambda_t=\frac{16(f\#)^2\nprimac\lambda_0}{16(f\#)^2\nprimac+n^2},
 \end{equation}
 and the blue shift $\Delta\lambda_0=\lambda_t-\lambda_0$ is then given by
 
 \begin{equation}
 \Delta\lambda_0=-\frac{\lambda_0}{16(f\#)^2+n^2}\frac{n^2}{\nprimac}.
 \end{equation}
 For large values of the f-number compared to the refraction index of the medium in which the etalon is immersed, we can simplify this expression to
 
 \begin{equation}
 \Delta\lambda_0 \simeq -\frac{\lambda_0}{16(f\#)^2}\frac{n^2}{\nprimac}.
 \end{equation}

\section{B:Transmitted electric field}
\label{appendix2}
If we denote by subindices 1,2,...$N$ the first, second and successive transmitted rays until the $N$th ray in Fig. (\ref{fig:Etalontransref}), their electric fields are given, following to the notation presented in Sect.\ref{sec:basicparameters}, by
\begin{equation}
\begin{gathered}
\Et_1=tt'{\rm e}^{{\rm i}\delta/2}{\bf E}^{\rm (i)},\\
\Et_2=tt'r^2{\rm e}^{{\rm i}(\delta/2+\delta)}{\bf E}^{\rm (i)},\\
\Et_3=tt'r^4{\rm e}^{{\rm i}(\delta/2+2\delta)}{\bf E}^{\rm (i)},\\
\vdots\\
\Et_N=tt'r^{2(N-1)}{\rm e}^{{\rm i}\left(\delta/2+(N-1)\delta\right)}{\bf E}^{\rm (i)},
\end{gathered}
\end{equation}
where a global phase $\delta/2$ has been included to take into account that the electric field is retarded in the first pass with respect one to the incident by the amount $2\pi \lambda^{-1} n\prima h \cos\theta\prima$. The transmitted electric field would be the superposition of each individual ray. Note that the transmitted rays follow a geometric sequence of common ratio $r^2{\rm e}^{{\rm i}\delta}$. As $r<1$, the sum of all rays can be expressed as

\begin{equation}
{\bf E}^{\rm (t)} = \frac{T{\rm e}^{{\rm i}\delta/2}}{1 - R \, {\rm e}^{{\rm i} \delta}} \, {\bf E}^{\rm (i)}.
\end{equation}

The global phase is usually neglected as it disappears when calculating the transmitted intensity by complex conjugating the electric field. This phase cannot be neglected for telecentric illumination of the etalon, as it depends on the incidence angle and, thus, on the pupil coordinates.



\end{document}